# Dynamics and Spatial Distribution of Global Nighttime Lights

**Nicola Pestalozzi, Peter Cauwels, and Didier Sornette**

ETH Zurich

Department of Management, Technology and Economics,

Chair of Entrepreneurial Risks, Scheuchzerstrasse 7, CH-8092 Zurich

nicola.pestalozzi@gmail.com, pcauwels@ethz.ch, dsornette@ethz.ch

**Abstract:** Using open source data, we observe the fascinating dynamics of nighttime light. Following a global economic regime shift, the planetary center of light can be seen moving eastwards at a pace of about 60 km per year. Introducing spatial light Gini coefficients, we find a universal pattern of human settlements across different countries and see a global centralization of light. Observing 160 different countries we document the expansion of developing countries, the growth of new agglomerations, the regression in countries suffering from demographic decline and the success of light pollution abatement programs in western countries.

## 1. Introduction

In the mid-1960s, the U.S. Air Force started a research project called the Defense Meteorological Satellite Program (DMSP[1]). The main purpose of this enterprise was to provide the military with daily information on worldwide cloud coverage. After scrutinizing the first results, it was discovered that, besides the initial goal of measuring cloud coverage, nighttime light was also very well captured by the sensors. When the system was declassified in 1972, and the data made publicly available, this unexpected, but very convenient, side-effect, gradually gained more interest from the scientific community.

[1] http://www.ngdc.noaa.gov/dmsp/dmsp.html (Last visited, February 2013)

Light emission is a quantity that can be measured instantaneously, objectively and systematically. This is in stark contrast to many widely used economic and demographic indicators, which are often based on estimates and censuses, such as the gross domestic product, energy consumption, population levels, or the degree of poverty. In the standard approaches, biases, time lags and inaccuracies are unavoidable and the comparison between different countries is often problematic. In contrast, nighttime lights are remotely sensed from one single satellite, with the same resolution, at the same time-of-day, in a systematic way, covering the surface of the whole planet.

Since 1992, the images have been systematically digitized and are now freely downloadable from the web (NGDC, 2012). This has opened a treasure of data, available at no cost, to the scientific community. As a consequence, in the past decade, a new research field has prospered using these observations of nighttime light and its dynamics in space and in time, aimed at studying economic activity, demographics, energy consumption, poverty and development, conflicts, urbanism and the environment.

We will discuss the data, the different products that are freely available on the web and our additional processing and enhancement tools in section 2. In section 3, we will review the existing research and academic literature. From our synthesis of the wide range of papers using nighttime images, a best-practice approach will be proposed.

This paper will further contribute to the existing research and literature with the following unique results. Firstly, in section 4.1, we will calculate the dynamics of our planet's mean center of light. It will be shown that, over the past two decades, this center of light has gradually shifted eastwards over a distance of roughly 1000 km. This result will be compared with a recent study of the McKinsey Global Institute (MGI, 2012), which calculated the dynamics of the global economic center of gravity based on GDP figures.

Next, in section 4.2, a spatial light Gini coefficient will be introduced. This will make a comparison possible of how evenly light is distributed in different countries. We will come to the fascinating conclusion that, if only relatively dense urbanized areas are considered, the Lorenz curves and Gini coefficients of light are almost identical in every country, even though the considered area can differ up to two orders of magnitude. Additionally, it will be shown that the Gini coefficients follow a slightly increasing trend over the past two decades. This is evidence of a continuous centralization of light, or, the gradual increase in the inequality of the spatial distribution of light.

Global light is centralizing. In section 4.3, we will dig deeper into this observation, analyzing the dynamics of bright versus dimly lit areas. It will be shown that dimly lit areas have increased proportionally more than bright areas, with a 49% compared to a 19% growth rate over the past 17 years. From this, it can be concluded that the area of dimly lit settlements has increased proportionally more than the area of the bright cities. Maybe counter-intuitively, this appears to be the major driver behind the observed increase in spatial light Gini coefficients over that period. Further, in this section, we will compare the nighttime light dynamics of economic tigers like China, India and Brazil, with countries that have seen a sensitive reduction in the urban population like Russia (excluding Moscow), Ukraine and Moldova. Additionally, it will be shown that, in some developed countries like Canada or the U.K., the size of the largest light agglomerations is slightly diminishing. This demonstrates the relative success of light pollution abatement programs launched in different countries around the world. Finally, the notion of a country will be dropped and the concept of light-defined agglomerations will be introduced. Tables and graphs will be given of the 20 largest light agglomerations worldwide in 1992 compared with 2009. This will demonstrate quantitatively how dramatically fast-growing megalopolis are changing in some developing countries like China and Egypt.

In this paper, we will look at the dynamics of global light growth covering almost two decades of available data. A review of the literature reveals that the majority of existing studies makes use only of single one-year snapshots and does not study time evolutionary processes. One of the reasons for this is the complex data handling and inter-calibration between the different satellite sensors. We therefore explain in details how we have addressed this problem.

## 2. Data and methods

### 2.1 The Satellites

In the DMSP program, typically two satellites are orbiting simultaneously, in a sun-synchronous low earth orbit, at an 833 km altitude. The satellites, which have a lifespan of about 6 to 8 years, pass over any given point on Earth between 8.30 pm and 9.30 pm, local time (Elvidge et al., 2001). They are equipped with a so-called Operational Linescan System (OLS)[2]. This consists of two sensors, one operating in the visible, near-infrared (400 to 1100 nm) spectrum and the other operating in the thermal infrared (10.5 to 12.6 μm) domain. Each detector has a field of view of 3000 km and captures images at a resolution of approximately 0.56 km.

### 2.2 The Stable Lights product

Creating good-quality scientific products from the raw data of the satellites is an undertaking of monumental difficulty and requires a huge processing effort. The National Geophysical Data Center (NGDC) of the National Oceanic and Atmospheric Administration (NOAA) has a research project dedicated exclusively to this task.

Firstly, the usable area of each image must be selected taking into account daylight scattering. Further, the parts with a scan angle greater than a certain threshold have to be excluded because these suffer from background noise and give a poor accuracy on the geo-location (Baugh et al., 2010). Then, the images are re-projected into 30 arc-second grids; this represents an area of about 0.86 square kilometers at the equator. A geo-location is assigned to each pixel of the usable area, based on different variables such as scan angles, satellite altitude and azimuth. In order to obtain a composite that covers the whole globe, all the selected satellite images are entwined. In the end, each composite represents a one-year average nighttime light image of the world, covering an area between 180 and -180 degrees longitude and -65 and 75 degrees latitude. Some parts of Greenland, Alaska, Canada, Scandinavia and Antarctica are missing. However, it has been estimated that only about 10,000 people or a mere 0.0002% of the world population lives there (Henderson et al., 2012). Since many observations were disturbed by clouds, moonlight, sun glare and other factors, it is estimated that each pixel is the result of 20 to 100 clear observations (Elvidge et al., 2009).

[2] http://www.ngdc.noaa.gov/dmsp/sensors/ols.html (Last visited, February 2013).

Exhibit 1 shows the list of available composites. When two satellites were orbiting at the same time, two different composites were produced. This creates a redundancy that can be used to inter-calibrate the images. We will come back to this in the following section. At the time of writing, there are 31 composite images available covering a time-span of 18 years. For each of these composites, a product called Stable Lights is available. In this product, fires and other ephemeral lights are removed, based on their high brightness and short duration. In the final result, each pixel quantizes the one-year average of stable light in a 6-bit data format. The pixel values, which we will call Digital Number (DN) in the following part of this paper, are integers ranging from 0 to 63. A comprehensive description of the methodology is given in (Baugh et al., 2010). It should be mentioned here that there are some artifacts in this final product, which have also been discussed by (Henderson et al., 2012). As a result of the noise and unstable light removal, the number of dim pixels, with a DN below 5, is unreasonably low, e.g. in the year 2000 composite of satellite F14 (see Exhibit 1 for an overview), there are no pixels with a Digital Number equal to 1. A second artifact is the occurrence of a saturation, which can be seen in the higher end of the spectrum for pixels with a DN of 60 and above. We have carefully considered these matters during the measurements and data analyses presented in this paper and will discuss them when relevant in our analyses presented below.

| Year | Satellites |
|---|---|
| 1992 | F10 |
| 1993 | F10 |
| 1994 | F10, F12 |
| 1995 | F12 |
| 1996 | F12 |
| 1997 | F12, F14 |
| 1998 | F12, F14 |
| 1999 | F12, F14 |
| 2000 | F14, F15 |
| 2001 | F14, F15 |
| 2002 | F14, F15 |
| 2003 | F14, F15 |
| 2004 | F15, F16 |
| 2005 | F15, F16 |
| 2006 | F15, F16 |
| 2007 | F15, F16 |
| 2008 | F16 |
| 2009 | F16 |
| 2010 | F18 |

**Exhibit1 –** The available composites. For most of the time, two satellites were orbiting simultaneously.

Following an attentive study of the literature and a consultation with the authors of the dataset, we decided that the Stable Lights product is the most suitable for our research. All the described methods and analyses from here onward refer therefore to the Stable Lights dataset, version 4 (NGDC, 2012).

## 2.3 Data Processing

### *2.3.1 Tools*

Dealing with geographical images of up to 3 Gigabytes in size is not trivial. Therefore, all the preliminary data processing is done using the ArcGIS[3] software package; this is the Geographic Information System from ESRI[4]. The country boundaries used in this research were downloaded as a polygonal shapefile from CShapes (Weidmann, et al., 2010). All countries that ceased to exist (e.g. Yugoslavia, USSR) were removed, such that a total of 194 non-overlapping countries remained in the dataset. For the geo-location of big cities, the point-based shapefile from (Nordpil, 2009) was used. It features the center of the 589 cities with a population larger than 750'000 in 2010. Additional analysis of the data was done using MATLAB[5].

### *2.3.2 Gas flares removal*

Gas flares are combustion devices used mainly in oil wells and big offshore platforms to burn flammable gas (mostly methane) released during the operations of oil extraction. They are a continuous phenomenon. Consequently, they are not filtered from the Stable Lights product. As the goal of this research is to study human settlements, they should be removed in order to avoid their misinterpretation as small cities. Gas flares have a characteristic circular shape of saturated pixels with glowing surroundings. These features make it possible to map their location (Elvidge, et al., 2009). The shapefiles, containing the gas flares' geo-location, one per country, were downloaded from the NGDC[6] website and then merged within the ArcGIS system. The obtained mask was then converted into a binary raster so that the gas flares' locations were given value zero, whereas all others pixels had a value of one. Every Stable Light composite was then multiplied with this raster to obtain images free of gas flares. Unfortunately, the polygons that encircle the gas flares are relatively large. Thus, it is unavoidable that certain areas of human-made lighting are improperly canceled out in this procedure. This small drawback, however, does not outweigh keeping the whole set of gas flares in the composites.

---

[3] ArcGIS Desktop 10.0, Service Pack 5.
[4] Environmental Systems Institute – http://www.esri.com (Last visited, February 2013).
[5] Matlab 7.12.0
[6] http://www.ngdc.noaa.gov/dmsp/interest/gas_flares_countries_shapefiles.html (Last visited, February 2013).

*2.3.3 Inter-calibration*

In this paper, we research the dynamics and the evolution in time of global nightlight, covering two decades of available data. The comparison between different image-years must be done with great care. Different satellites have different sensor settings. Also for the same satellite, the captured images are subject to a natural deterioration of the sensor over time and of undocumented gain adjustments.

To correct for these effects, we decided to apply the method of Elvidge (Elvidge, et al., 2009), because it is fully documented, complete and has been used as an academic standard in different previous studies (e.g. (Zhao, et al., 2010), (Small, et al., 2010), (Small, et al., 2012), (Chen, et al., 2011) and (Elvidge, et al., 2011)). Three calibrating parameters, $C_0$, $C_1$ and $C_2$, are calculated for each year and satellite. In this way, the new, inter-calibrated, Digital Numbers (DN) can be calculated directly from the old numbers, for each image, using the following equation:

(1) $DNnew = C0 + C1 * DN_{old} + C2 * DN_{old}\hat{}2$

The inter-calibration coefficients that are provided in the referenced publication only cover the years 1994 until 2008. However, additional calibration coefficients were found, for the year 2009, in a not yet published paper by the same authors (Elvidge, et al., 2011). In the end, only the 2010 image of the satellite F18 had to be removed from the analysis because no calibration coefficients were obtainable.

An indicator of the goodness of the calibration is that the Sum of Lights (SOL), i.e. the sum of all pixel values for a certain region, matches between two composites of the same year coming from different satellites. Exhibit 2 gives the example of China. It can be clearly seen that the overlaps between the different satellites is much smoother after calibration.

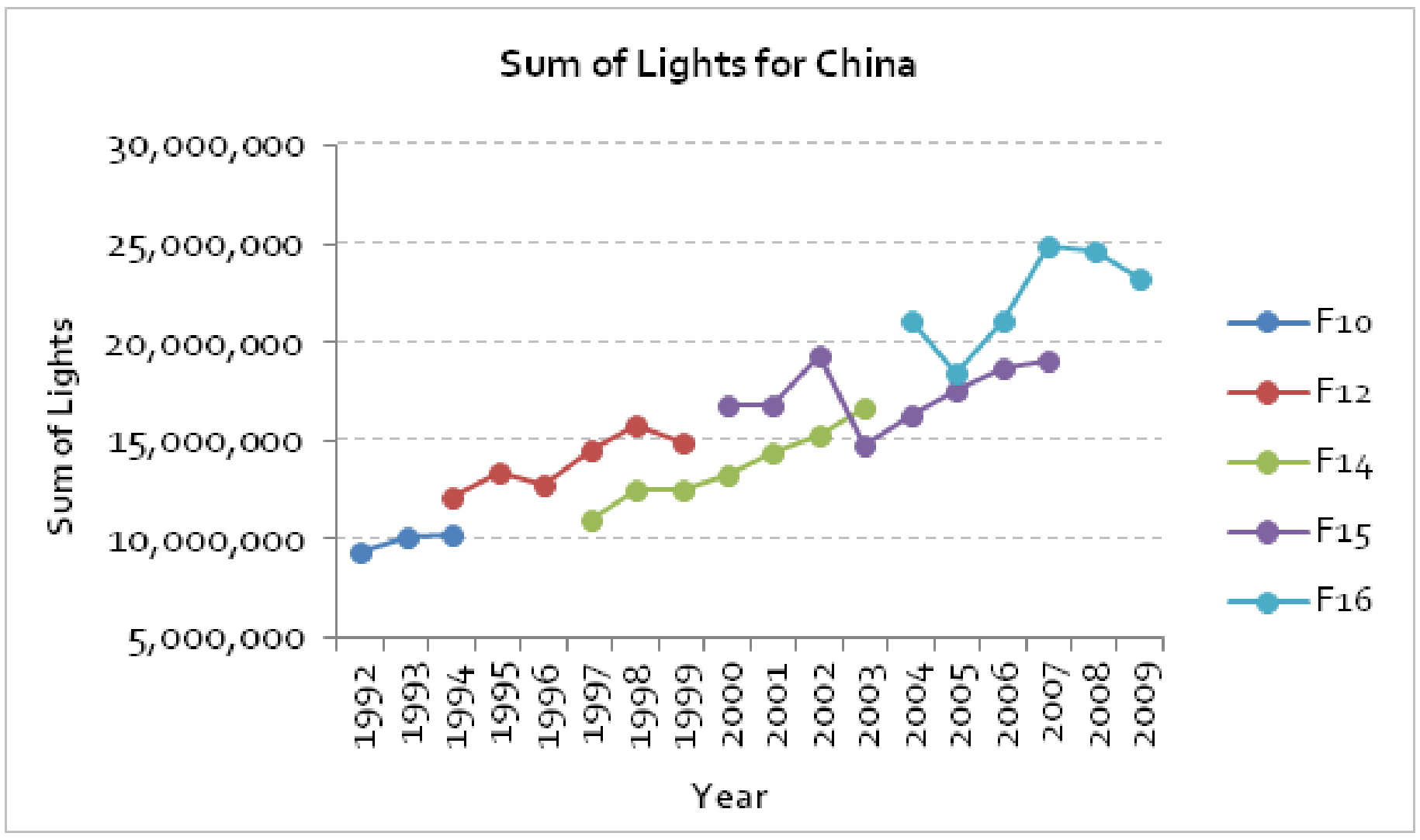

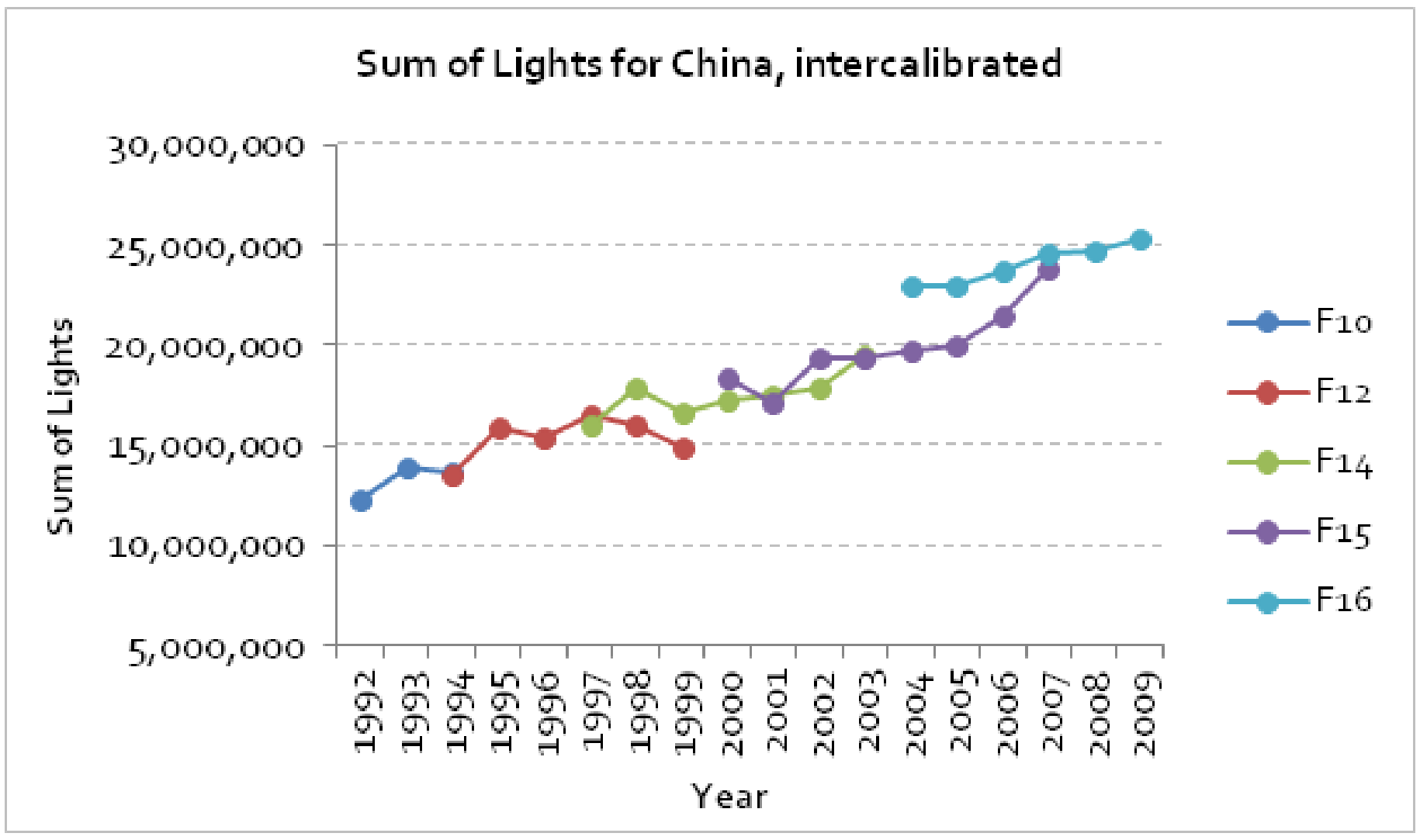

**Exhibit2 –** Sum of lights for China before (top) and after (bottom) the inter-calibration.

#### *2.3.4 Re-projection*

The 30 arc-seconds grid, used in the composite, does not correspond to an equal area at the surface of the earth. For example, in Quito (on the Equator), one pixel tallies a 0.86 square kilometer surface, whereas in Reykjavik (at 64°N latitude), it fits a surface less than half this size, or 0.41 square kilometer. Since we want to analyze spatial dynamics of nightlight, a re-projection is needed so that each new pixel or cell in the satellite image corresponds to an equal area at the planet's surface. For this purpose, the Mollweide projection was applied. This is a pseudo-cylindrical map projection where the accurate representation of areas takes precedence over the shape.

## 3. Literature review

The availability of the digitized and freely downloadable nighttime light images from the NGDC website is a treasure of data that has instigated new research in economics, social sciences and environmental studies. Nighttime lights can be remotely sensed, objectively, at the same time-of-day, in a systematic way, covering the surface of the whole planet. This is in stark contrast to many widely used economic and demographic indicators, which are often based on estimates and censuses, where biases, time lags and inaccuracies are unavoidable and the comparison between different countries is problematic.

### 3.1 Economics

Nighttime lights provide an appealing innovative instrument to measure economic activity. However, the relationship between economics and light is not entirely trivial, which makes it difficult to construct reliable estimators. Nevertheless, especially for countries with poor data quality, the approach can add significant value to existing statistics.

In one of the first economic studies applying nighttime light data, Doll et al. (2005) used the 1996 night images of 11 European countries. They found a very strong linear relationship (with an $R^2$ value of 0.98) between the night light energy emission of these countries and their nominal GDP. This result was further confirmed by Sutton (2007). Using the Stable Light product for the year 2000 of the U.S., India, China and Turkey, they found a log-linear relationship between the night light energy emission and the nominal GDP (with an $R^2$ value of 0.74). Similar studies were conducted for China (Zhao, et al., 2010), India (Bhandari & Roychowdhury, 2011) and Mexico (Ghosh, et al., 2009), always using a one-year snapshot only.

In more recent work, Chen & Nordhaus (2011) and Kulkarni et al. (2011) link the time evolution of nighttime light to economic activity. They do not limit their research to one specific year only, but make use of all of the available data. They conclude that light can be used as a proxy for nominal GDP, but that this approach only adds value to the official statistics for countries with poor reporting standards and low data quality. This was further confirmed for GDP growth by Henderson, et al. (2012), who found that growth estimates, based on nighttime light images, differ substantially from the official statistics for countries with low data quality. A case in point is Myanmar. According to the World Development Indicator of the World Bank (WDI), the average annual growth rate, between 1992 and 2006, was 10%. This is to be compared with a 3.3% estimate based on the nighttime light images. For Burundi, on the other hand, the WDI predicted a GDP decline of -0.3%, whereas satellite data implied a growth of 2.9% annually, over that same period.

### 3.2 Demographics and politics

Persistent light is a clear indicator of the presence of human settlements. An early attempt to estimate population densities with nighttime light images was done by Sutton (1997). He compared data from the 1990 U.S. census with a binary image containing only the saturated pixels. It was found that these images could only explain 25% of the variation in the population density of the urban areas in the continental U.S. In a later study, the same author and collaborators estimated the global human population for the year 1997. He came up with a figure of 6.3 billion people compared to 5.9 billion, which was the generally accepted estimate for that year (Sutton, et al., 2001). Similar studies were done at the national levels, e.g. for China (Lo, 2001; Zhuo, et al., 2009) and Brazil (Amaral, et al., 2006). Light alone may not be a perfect proxy to measure populations but, combined with other sources, it can substantially add value. It is worth noting that the Gridded Population of the World (GPW)[7], which is one of the most widespread databases of the global population density, uses nighttime lights as one of its many inputs.

Also using nighttime light images, Elvidge, et al. (2009) constructed a global Poverty Index (PI) by dividing population numbers by the average light. This shows the regions where people are living without satellite detectable stable light, which is, according to the authors, an indicator for poverty. After calibrating the index with official statistics on poverty, it was estimated that 2.2 billion people live with an income below $2 a day. This is to be compared with the 2.6 billion estimated by the World Bank.

Another application of nighttime light images is the mapping and measurement of urban boundaries. This was done by Imhoff, et al. (1997) for urban land areas in the U.S and later by Henderson, et al. (2003) for three distinct cities: Lhasa, San Francisco and Beijing. In that same field, Small et al. (2005) came up with the very interesting result that conurbations, with a diameter larger than 80km, account for less than 1% of all settlements but for about half the total lighted area worldwide. Other detailed studies on the level of urbanization in China can be found in (Elvidge, et al., 2007; Lo, 2002; Ma, et al., 2012).

[7] http://sedac.ciesin.columbia.edu/data/collection/gpw-v3 (Last visited, February 2013).

Some authors suggested that conflict related events also could have an impact on the light emission of a region and may, as such, be studied by means of nighttime images. One such study evaluates the U.S. military surge in Iraq in 2007 (Agnew, et al., 2008), which was aimed to stabilize and rebuild the cities after the war. The authors expected an increase in luminosity over time, because of the restoration of the electrical infrastructure. However, no observable effect could be found. Another study looked for effects of the wars in the Caucasus regions of Russia and Georgia (Witmer & Loughlin, 2011). The authors claim to be able to detect oil fires and large refugee outflows, as well as settlement (re)construction. In summary, the one-year time resolution of the images may be too coarse to detect the actual impact of wars, but for large conflicts with a longer duration, light can be used to study the impact of damages and the dynamics of reconstruction.

The politics of electricity distribution in certain regions of India was studied by Min (2009). Although there is a relatively reliable network infrastructure, the actual power supply is by far not sufficient to meet the demand. The decisional power of giving energy to a certain municipality, district or county is quite centralized and involves political control. However, only a weak correlation could be found between the availability of energy in the municipalities and the political party that won the elections.

### 3.3 Environmental studies

One obvious application of the night light data is to measure and estimate energy consumption. In an early study, Elvidge et al. (1997) showed a very strong log-log relationship (with an $R^2$ value of 0.96) between the lighted area and the energy consumption at a country level. Further similar studies were conducted at the national level for India (Kiran Chand, et al., 2009), Brazil (Amaral, et al., 2005) and Japan (Letu, et al., 2010).

As mentioned earlier in section 2.3, other, probably less known, features recorded in the light maps are the so-called gas flares. These are combustion devices to burn flammable gas released during the operations of oil extraction. The gas is burned only for convenience, or for lack of infrastructure, so that the energy is wasted and huge amounts of $CO_2$ are released in the atmosphere. The monitoring of this harmful and quite uncontrolled activity is motivated by environmental and health concerns, besides energy efficiency reasons. Most gas flares burn uninterruptedly. Consequently, this phenomenon is still observable in the Stable Lights products. Using night light images, Elvidge, et al. (2009) estimated that, in 2008, approximately 139 billion cubic meters gas were wasted on a global scale. This is equivalent to 21% of the total natural gas consumption of the U.S. and has a retail market value of $68 billion and an impact on the atmosphere of 278 million metric tons of $CO_2$ equivalent.

Large forest fires (natural and human made) can also be visible from space. Studies were done to monitor the surface of forests affected by fires in India (Kiran Chand, et al., 2007), Indonesia (Fuller & Fulk, 2010) and Brazil (Elvidge, et al., 2010).

Even some fishing activities can be seen from outer space. This is the case for a specific technique where bright lights are mounted on the boats to attract squid. The illumination can be so intense and persistent that it remains visible after the filtering procedure in the Stable Lights images. Particular studies on this subject were made for Japan (Kiyofuji & Saitoh, 2004), where the spatial and temporal variability of night fishing were analyzed to better understand the migration of squid. The impact on ecological systems around coral reefs, where the illumination is seen as a stressor and a threat, was studied by Aubrecht, et al. (2008).

There are many reasons why light pollution can have negative consequences: firstly, astronomical light pollution reduces the number of visible stars and disturbs the scientific observation of the sky. Secondly, the ecological light pollution represents a threat to entire ecosystems, substantially altering the behavioral patterns of the animal population (orientation, foraging, reproduction, migration, communication and so on.). And finally, wasted light means also wasted energy. Some human health disorders were also found to be correlated with prolonged exposure to light during night. Studies on light pollution, using the nighttime light images, have been done for Europe (Cinzano, et al., 2000; Cinzano & Elvidge, 2004) and Iran (Tavoosi, et al., 2009).

## 3.4 Discussion

Making the nighttime light images available for the scientific community has instigated a lot of research covering a broad range of disciplines. Exhibit A1 in the appendix 6.1 summarizes all the publications that were cited in the preceding literature review. This table makes it possible to do a comparative and a qualitative analysis between the different studies and may be helpful in setting up a best-practice in using the night images for different research purposes.

It can be seen that most of the studies used a single one-year snapshot only, although the images are available for 18 satellite-years. More surprisingly, however, is the fact that quite a few of the analyses that cover multiple years of data did not perform an inter-calibration of the images. It is known and documented (Elvidge, et al., 2009), however, that the sensors have different sensitivities (for example, F14 produced substantially dimmer images (Doll, 2008)), and even for the same satellite, in addition to the natural deterioration of the sensor over time, undocumented gain adjustments were made during its lifetime, so that the comparison between different image-years is really delicate.

For most of the studies that were subject to this review, the gas flares are not relevant; they should be removed before performing any analysis. However, this operation was hardly ever done, or else, it was not mentioned explicitly in the publication. This phenomenon is present in at least 20 countries, notably Russia, Nigeria, Iran, Iraq and Algeria, and, especially for small countries (such as Qatar, Kuwait) it could lead to a misinterpretation of the results, since gas flares can be easily confused with small cities. It was calculated for the year 2000, that gas flares represented 3.2 percent of the worldwide light emanation, and in some regions, such as sub-Saharan Africa (e.g. Nigeria) they accounted for even 30% of the total illumination (Henderson, et al., 2012).

Another very important preprocessing operation is the re-projection of the images with an equal-area method. This is always needed when analyzing spatial extent, because, as we explained in section 2.3.4, the pixels in the 30 arc-seconds grid represent different land areas depending on the latitude. Nevertheless, as can be seen in the table in Appendix, this process seems to be often neglected, or, not specifically mentioned in the publication.

Finally, the different products are also available in different versions, depending on the algorithm that was used for creating the Stable Lights image. It can be seen in Exhibit A1 that most publications fail to mention explicitly the specific version. This may complicate the reproduction and comparison of different studies.

In this research, we tried to follow a best-practice by inter-calibrating and re-projecting the different images and removing the gas flares. The Stable Lights dataset version 4 (NGDC, 2012) was used.

## 4. Results

### 4.1 The planet's mean center of light

In a recent study by the McKinsey Global Institute (MGI, 2012), called "Urban World: Cities and the rise of the consuming class", a method is presented to calculate the economic center of gravity of the world. The research uses data and approximations from the year 1 CE until 2010. Additionally, projections are given until 2025. The result shows that, propelled by the industrialization and urbanization of Europe and the U.S., the economic center of gravity (located near Kabul at year 1) has been gradually shifting towards the northwest (up to the vicinity of Reykjavik) until around 1950. After that, driven by the rise of Japan, the regime shifted and the dynamic reversed. Since then, in the last 60 years, the economic center of gravity has been moving back eastwards rapidly. The most recent decade, between 2000 and 2010, has seen the fastest rate of change, in global economic balance, in history. During this period, the world's economic center of gravity has shifted by about 140 kilometers per annum. Exhibit 3 graphically demonstrates the result of this study.

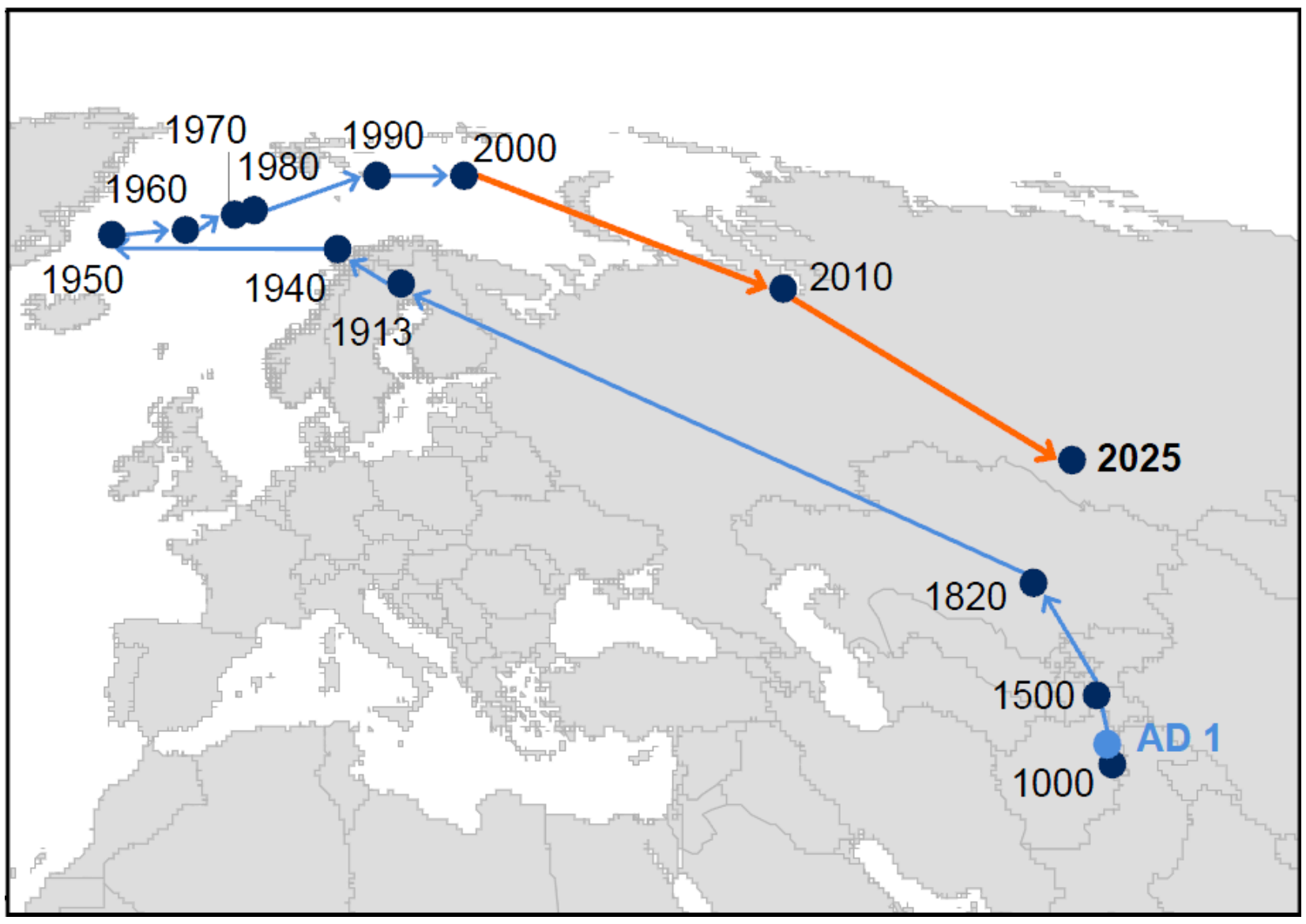

**Exhibit3 –** The dynamic of the global center of gravity (GDP based) from year 1 to 2010, and the projection for 2025. Source: (MGI, 2012)

The inter-calibrated nighttime light images that are at our disposal cover the period between 1992 and 2009. This includes exactly the period of the dramatic regime shift that can be seen in the MGI study. Therefore, we decided to challenge this fascinating study and compare the dynamics of the economic center of gravity with the center of gravity of nighttime light emissions.

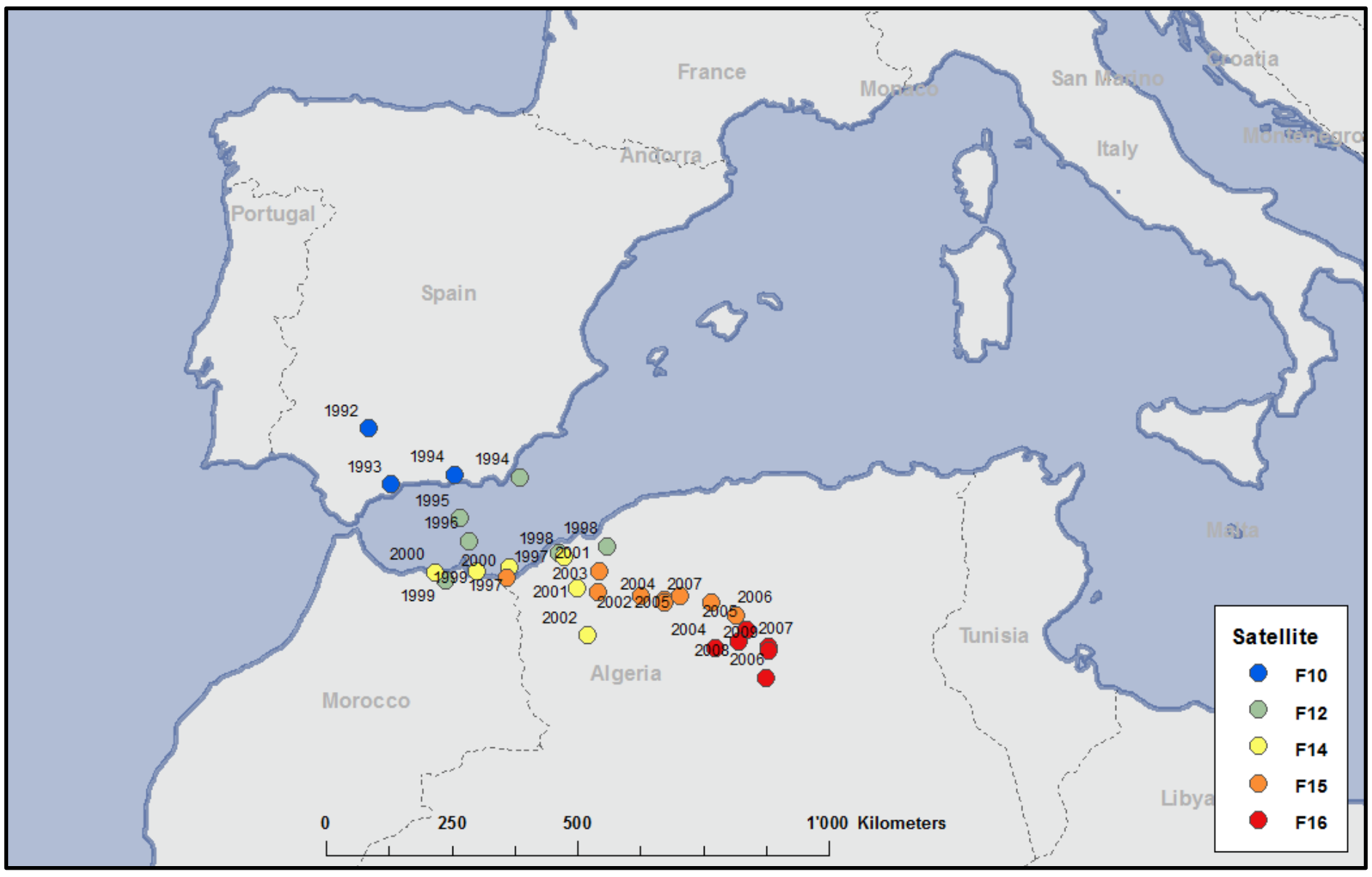

**Exhibit4** – The dynamic of the light mean center for the different satellites and years. Note the strong shift eastwards. The amplitude of the movement, between 1992 and 2009, is about 1’000 km; this is roughly half the estimate of MGI.

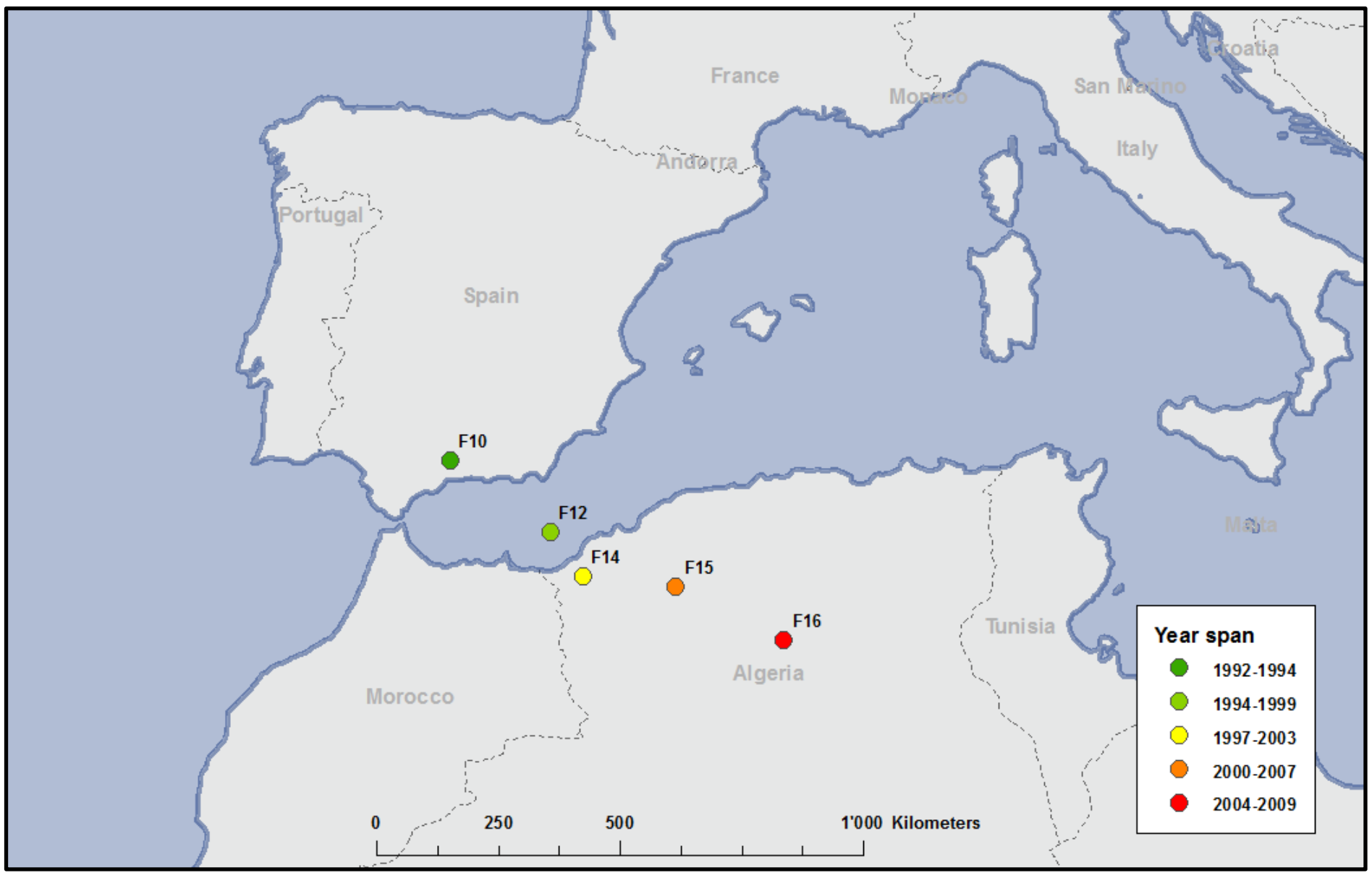

**Exhibit5 –** The dynamic of the light mean center for the different satellites. Here, the centroids are calculated on the average sum of lights per satellite (labeled), in order to emphasize the direction of the shift. Each satellite covers the year span indicated in the legend.

Exhibit 4 gives the results per satellite and year; Exhibit 5 shows the result per satellite over the years of its operation. The first fact is that the location of the light center deviates substantially from the economic center. Although the distinction could be partly explained by the use of different projection methods[8], this demonstrates that the comparison of light and economic indicators is non-trivial. The direction of the movement, however, appears to be consistent: there is a clear and strong southeastern shift, though the speed of this move is much smaller for the light center than for the economic center. The amplitude of the light movement is about 1000 km between 1992 and 2009; this corresponds to about 60 km per year, which is less than half of the 140 km from the McKinsey study.

## 4.2 The spatial light Gini coefficient

In this section, we analyze the distribution of nighttime light for different countries. More specifically, we want to understand the level of "centralization" of light. For this purpose, we will calculate spatial light Gini coefficients. These measure, in one single number, how the light is distributed over the area of a country. The lowest value of 0 corresponds to a complete equality, whereas the highest value of 1 is indicative for a total inequality; the higher the value, the more centralized the spatial distribution of light is. In our understanding, there are three distinct reasons why a country A would have a higher spatial light Gini coefficient than a country B, or, why the spatial light distribution in country A is more centralized:

- In country A, the small settlements occupy a larger fraction of the surface than the bright cities; the country area with low pixel values is relatively higher than the area with high pixel values;

[8] In this study, the "Mean Center" function of ArcGIS is used on the maps, after the Mollweide projection was applied. More details on the McKinsey Global Institute's approach can be found in their publication (MGI, 2012).

- In country A, the light emission of the cities is brighter; this is difficult to measure, however, because of the saturation effect of the satellite sensors;
- In country A, there is a lower light emission in the small settlements.

Often, in reality, the reason may as well be a combination of these three.

Exhibit 6 gives the Lorenz curves and the Gini coefficients for a selected group of countries. It can be seen that the coefficients cover a wide range from a value of 0.34 for the Netherlands to 0.98 for Brazil. In these calculations, the full range of pixel values in the images was used, including the ones with a Digital Number (DN) equal to zero, which correspond to areas where there is no light emission. However, in most cases, this means that these calculations also take into account the portion of land that is inhabitable. Consequently, the huge difference between the Netherlands and Brazil has mainly a geographical explanation: a large portion of Brazil is covered by the Amazonian rainforest whereas most of the Netherlands is habitable.

In Exhibit 7, we corrected for this by excluding the zero-pixels from the calculations. Thus, the Gini coefficients are calculated for land area that has a DN greater than, or equal to, 1. As such, we have removed all the unlit land from the analysis. The results are much different now, and the gap between the countries has, to a large extent, disappeared with coefficients ranging from 0.29 to 0.40. This is quite astonishing, especially if we compare the areas: the U.S has a lit area more than 70 times that of the Netherlands, yet, their Lorenz curves and the Gini coefficients are very similar. We take this approach a step further and gradually increase the lower cutoff of the Digital Numbers. The result is presented in Exhibit 8 and 9. It can be seen that the Lorenz curves and the Gini coefficients quickly converge for every country. This result suggests that the spatial configuration of the settlements follows a universal pattern across these different countries, even though the considered area can differ up to two orders of magnitude.

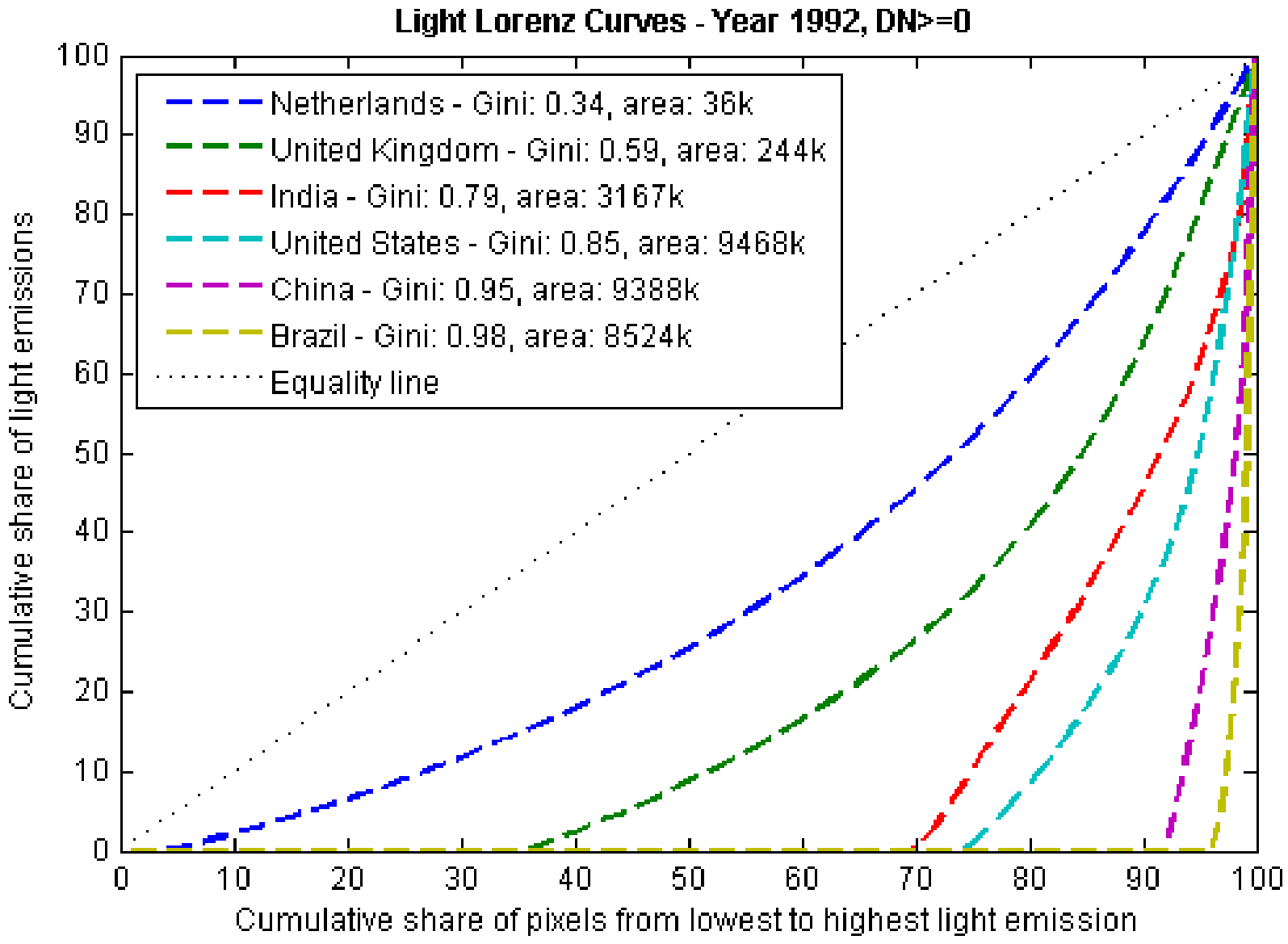

**Exhibit6** – The Lorenz curves and Gini coefficients for selected countries in 1992. The variable 'area' indicates the amount of square kilometers (in thousands) in the country corresponding with pixels that have a DN higher than the lower cutoff value (in this case DN>=0).

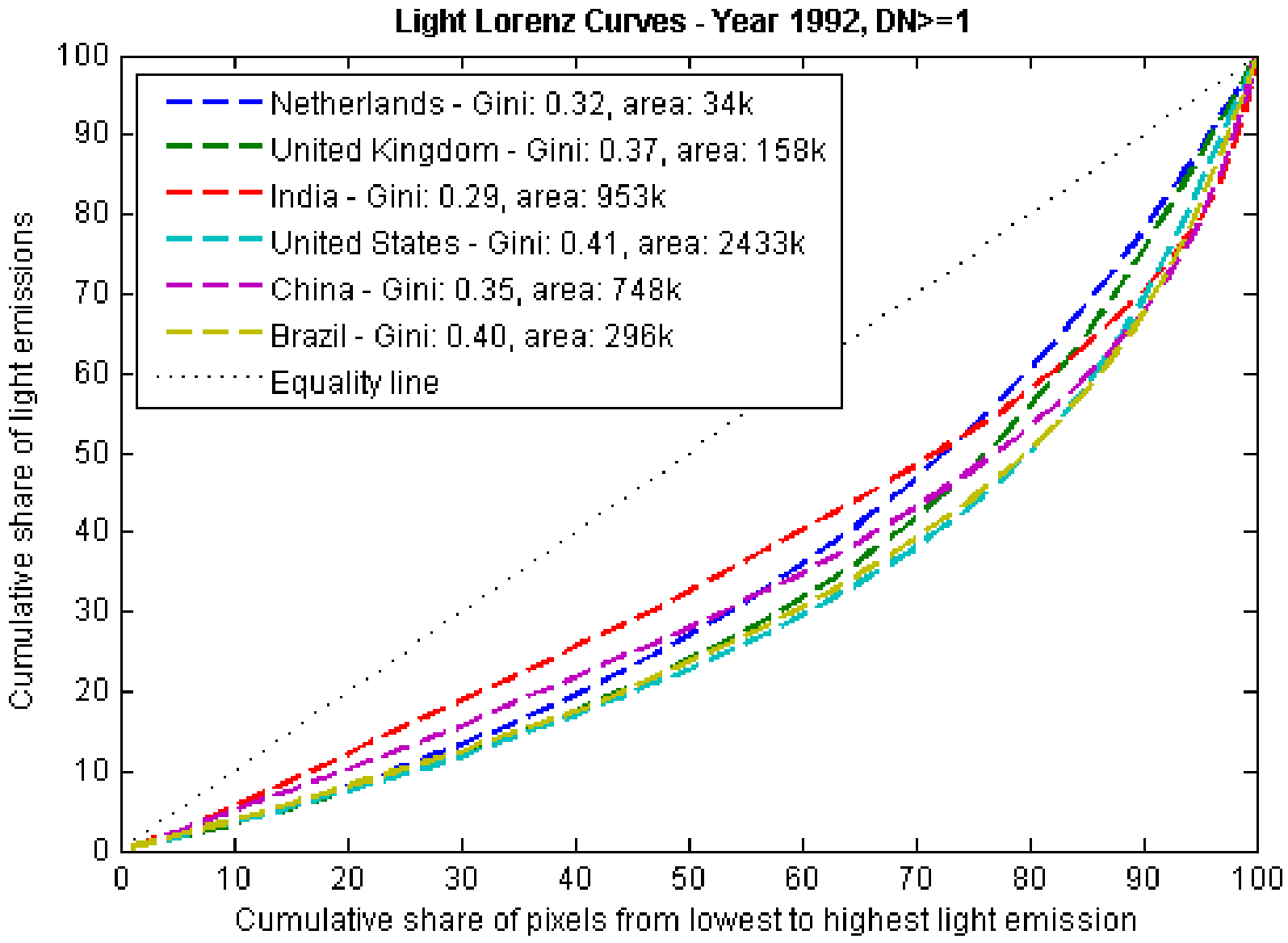

**Exhibit7** – The Lorenz curves and Gini coefficients for selected countries in 1992, with a threshold pixel value greater than or equal to 1. The variable 'area' indicates the amount of square kilometers (in thousands) in the country corresponding with pixels that have a DN higher than the lower cutoff value.

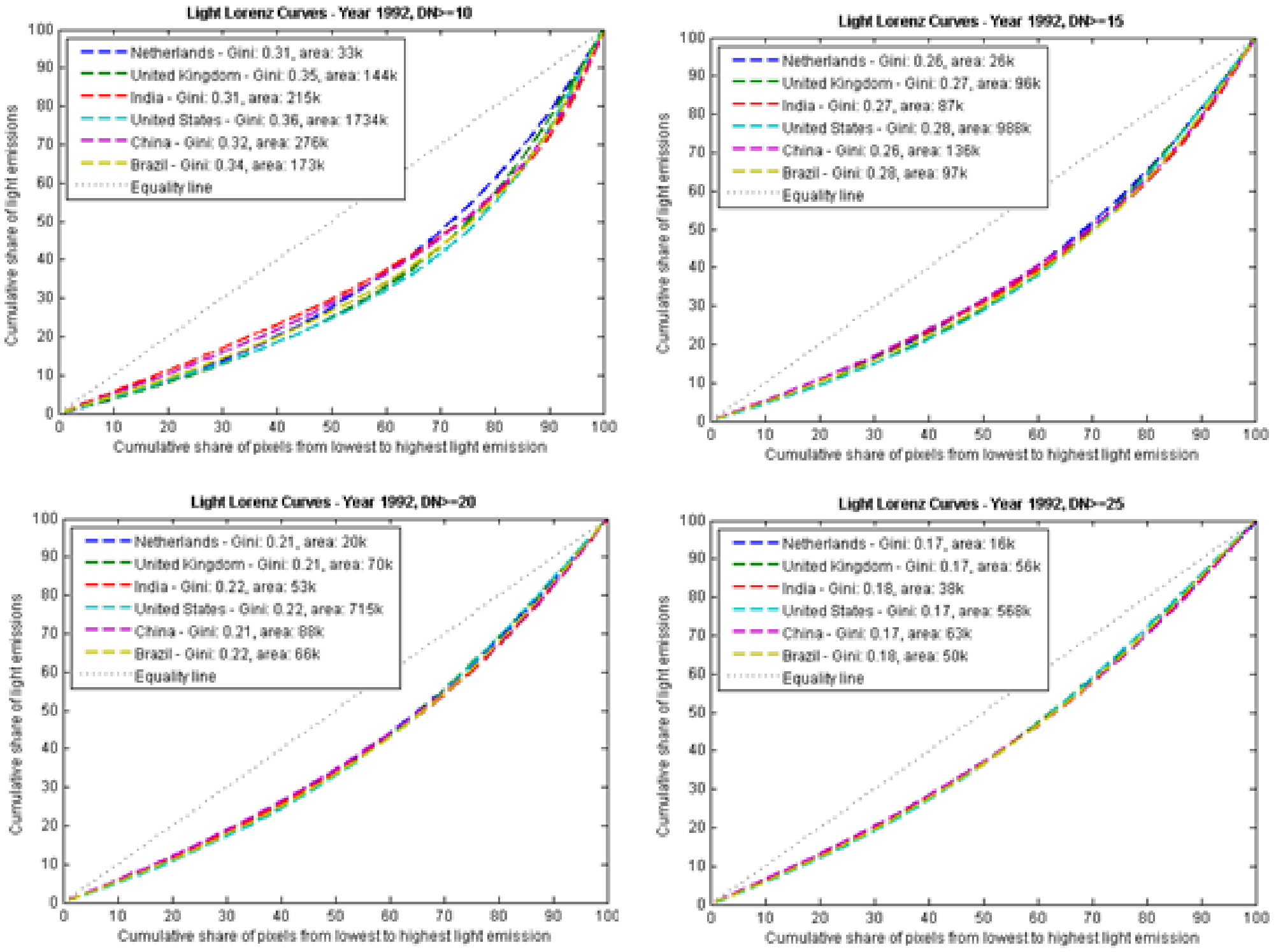

**Exhibit8 –** The Lorenz curves and Gini coefficients for selected countries in 1992, with increasing thresholds. Note how the Lorenz curves and the Gini coefficients converge. At the threshold of DN>=25, i.e. still far below the middle of the spectrum, the Netherlands and United States have the same Gini coefficient, although the considered lit area differs by a factor of 35.

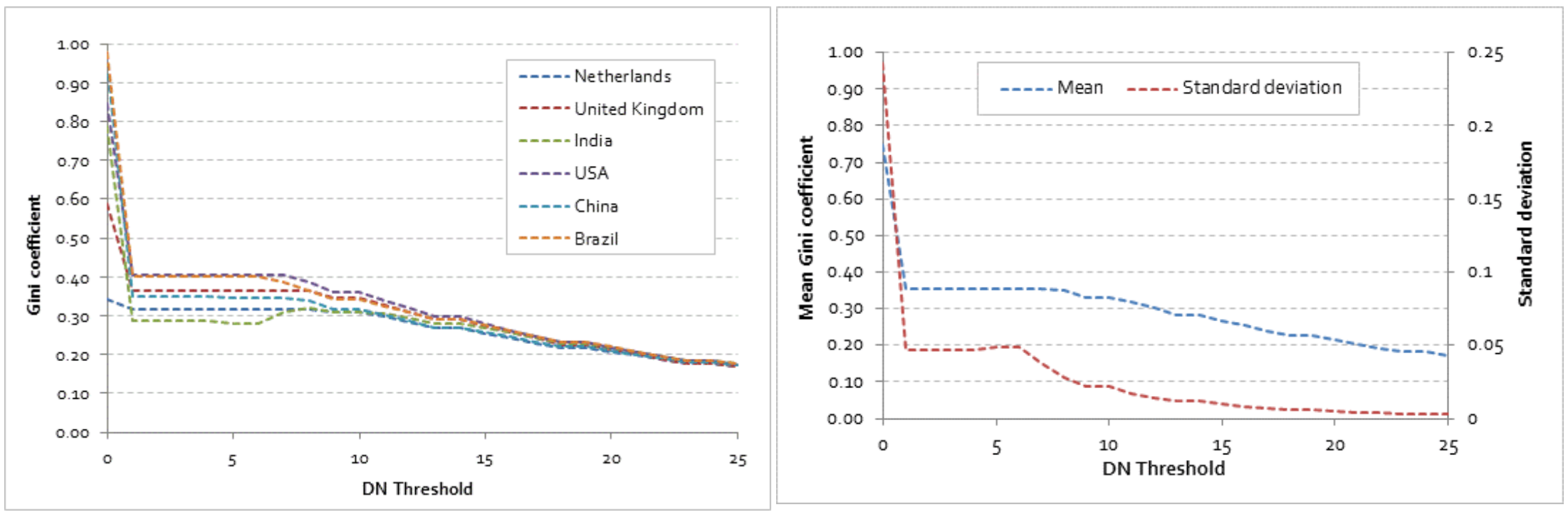

**Exhibit9 –** The change in the Gini coefficient for selected countries in 1992 using increasing thresholds.

This part of the analysis has mainly focused on a one-year snapshot. Let us now take a look at the historical evolution of the spatial light Gini coefficients for the selected countries. The left graph in Exhibit 10 gives the result without using any pixel value threshold. The graphs are basically flat, suggesting that no dynamical change in the distribution of nighttime light occurred. This is not a surprise because the result is dominated by the non-habitable land in the analysis. When we filter the images, however, and only take those pixels into account which have a DN higher than 4, the result is quite different. Now, a clear rise can be seen for all countries. This is presented in the right graph of Exhibit 10.

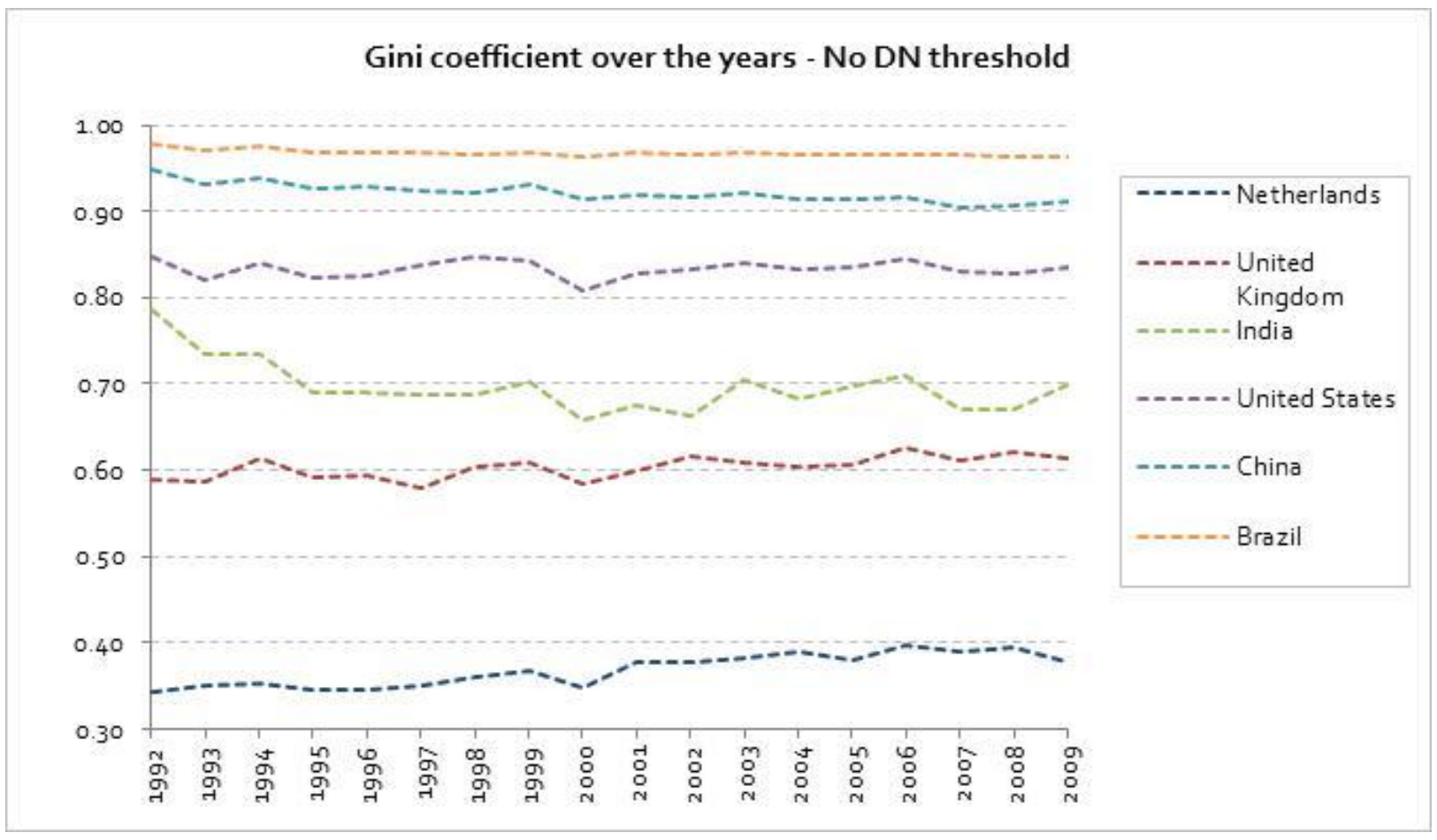

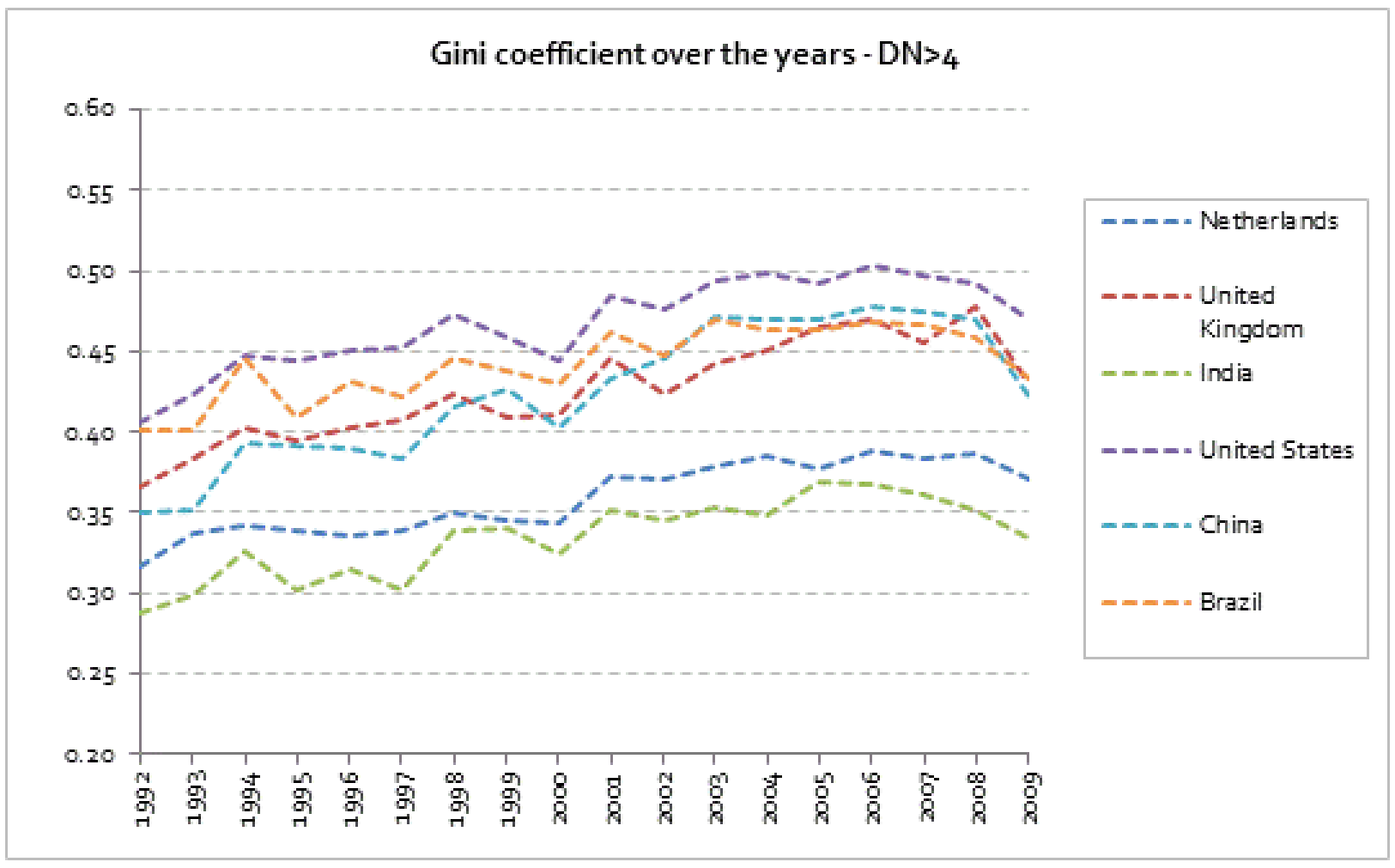

**Exhibit10 –** The historical evolution of the Gini coefficients for selected countries without any filtering (up) and using a threshold of DN>4 (down).

This suggests that spatial light has gradually centralized over the past 17 years. This can be explained as follows:

- The dimly lit area increased proportionally more than the brighter area. This suggests that villages and small settlements have been built (respectively electrified) over a surface greater than the one occupied by the city growth;
- The brightness of the cities increased proportionally more than the brightness of the small settlements. Assuming, for example, that the illuminated area did not change, in order for the Gini coefficient to rise, the brighter pixels have to increase their DN proportionally more than the dimmer ones;
- The overall brightness of the small settlements decreased, so that the relative share of luminosity taken by the bright lit agglomerations increased, driving the Gini coefficient up.

### 4.3 Bright versus dimly lit areas

We have come to the conclusion that spatial light Gini coefficients have gradually increased over the timeline of our observations, which is the past 17 years. Now, we will further analyze the process behind this observation. It is important to mention that, because of the pixel saturation in bright areas, it is not possible to use the nighttime light images to study changes in brightness. Thus, given the limitation of the tools at our disposal, we will have to focus on the change and evolution in the surface size of bright versus dimly lit areas.

In order to separate bright from dimly lit areas, binary images were produced using a threshold technique. Any pixel with a Digital Number higher than 1 and less than or equal to 30 was classified as dim; any pixel with a DN higher than 30 was classified as bright. This was done for the global image of the two individual years 1992 and 2009. The result of this exercise, for a total of 160 countries[9], is given in an extensive table in Exhibit 2A in the Appendix. A small extract from this table, showing the results for the BRIC countries, the U.S. and the World is summarized in Exhibit 11. It can be seen that between 1992 and 2009, the worldwide dimly lit surface grew 49%, whereas the bright area expanded 19%. From the table, it becomes clear that the area of dimly lit settlements has, both in relative as in absolute terms, increased more than the area of the bright cities. It may be counter-intuitive, but this is the major driver behind the observed increase in spatial light Gini coefficients over that period.

| | **Change of dimly lit area** | | **Change of bright area** | |
|---|---|---|---|---|
| | **Absolute** ($km^2$) | **Percent** | **Absolute** ($km^2$) | **Percent** |
| China | +574'609 | +81.41% | +95'394 | +225.76% |
| India | +448'852 | +48.41% | +21'380 | +80.95% |
| Brazil | +243'648 | +94.19% | +21'232 | +57.86 |
| Russia | +261'858 | +36.33% | -25'231 | -24.85% |
| USA | +511'651 | +81.41% | -6'316 | -1.44% |
| **World** | **+4'543'889** | **+48.92%** | **+261'530** | **+19.43%** |

**Exhibit11 –** Change in the lit area in the world and in selected countries between 1992 and 2009. A pixel threshold of 30 is applied to classify dim versus bright pixels.

To give a more dynamical view of the process observed in the table, Exhibit 12 shows the evolution of the total bright area, in square kilometers, for the four countries with the largest increase between 1992 and 2009. Three of the four BRIC countries are dominating the ranking, China outclassing the other two with an impressive absolute (95'394 km2) and relative (225%) growth. A clear concentration in the area around Shanghai and Shenzhen-Guangzhou could be seen. To show this, we generated images, which can be found in the Appendix 6.3. The high ranking of Egypt may come as a surprise. This is, however, the result of the remarkable development of the delta and the coast along the Nile. For the images, demonstrating these assertions, we refer to the Exhibits A3-A5 in the Appendix. It is also worthwhile mentioning that the evolutions of India and Brazil have been more diffuse. No clearly defined hot spots could be observed during the analysis, as was the case for China and Egypt.

[9] Countries with an area of less than 5'000 square kilometres (mainly islands) were excluded because of misleading results in the percentages. A total of 160 countries remained in the database.

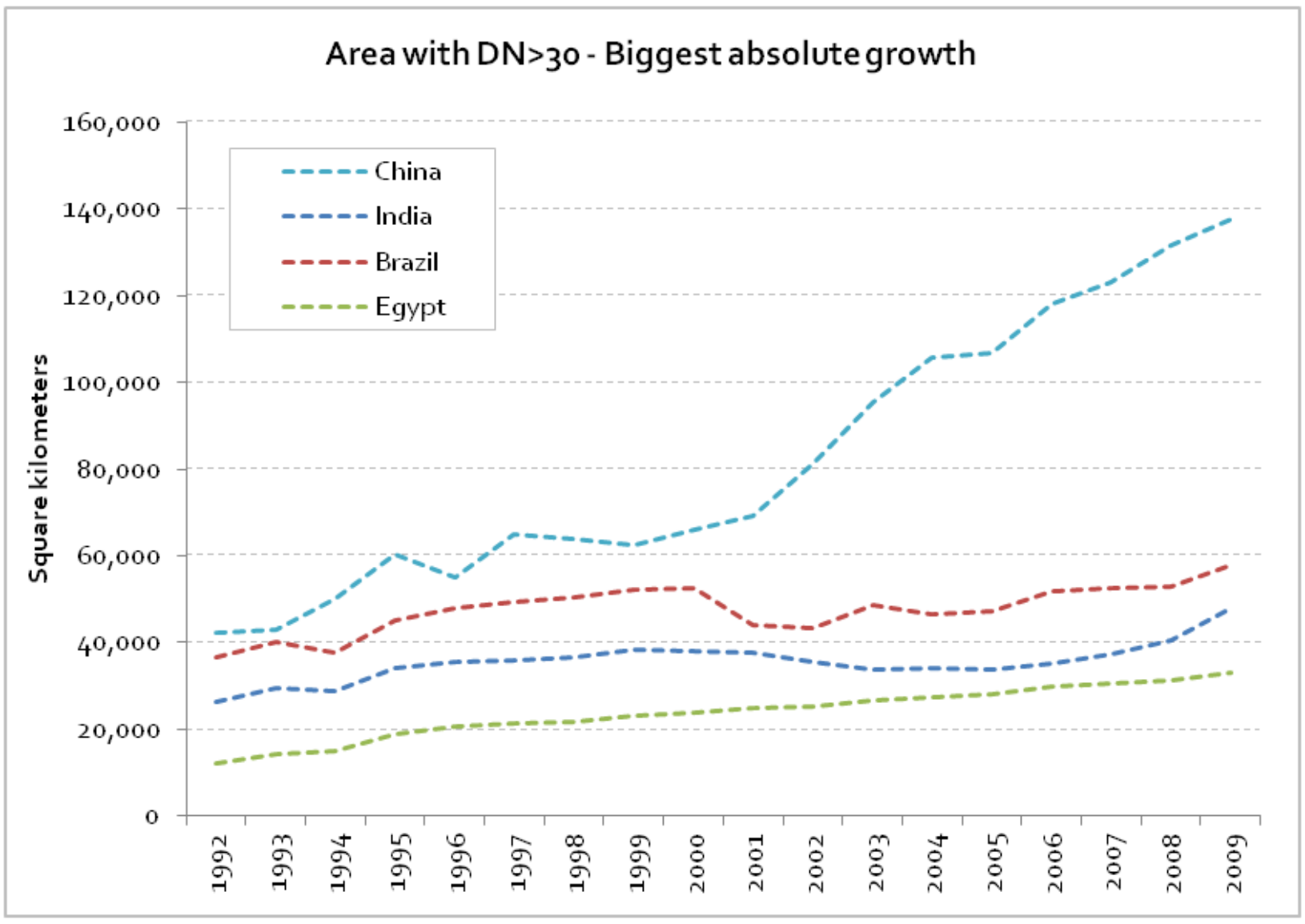

**Exhibit12** – Evolution of the area with DN>30 for the countries with the highest absolute increase between 1992 and 2009.

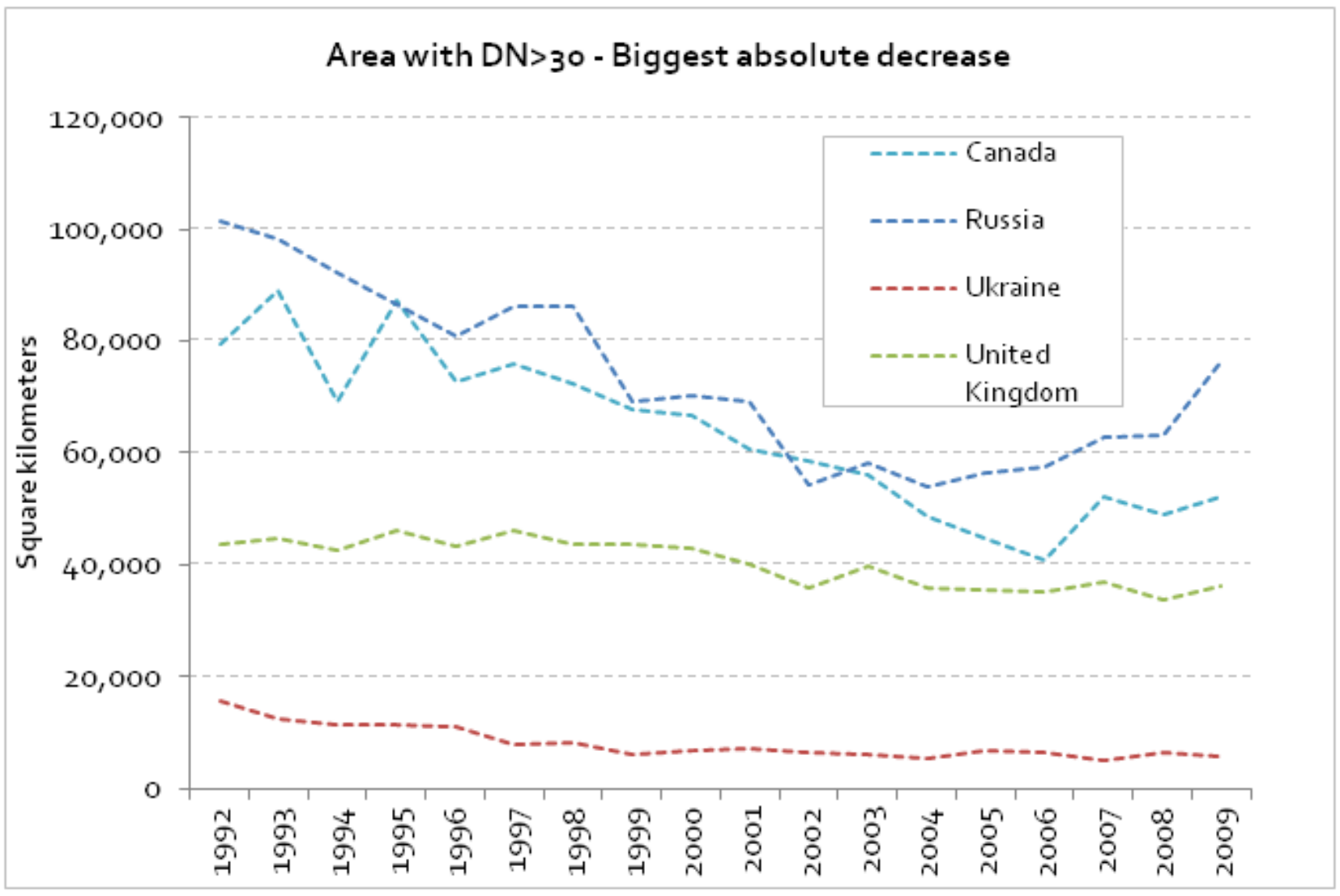

**Exhibit13** – Evolution of the area with DN>30 for the countries with the largest absolute decrease between 1992 and 2009.

Not all countries have seen an increase in bright area. A different dynamic is demonstrated in Exhibit 13. The graph shows the four countries with the largest absolute decrease. We can see two different processes at work here. For Russia and the Ukraine, the absolute decrease in the bright area goes hand in hand with a decrease in the urban population. This is further demonstrated in the table in Exhibit 14, which gives an overview of the countries where the decrease of bright areas coincided with a considerable reduction of the urban population. In Canada and the United Kingdom, on the other hand, another process is at work. For those countries, we conjecture that the decrease is the result of the Light Pollution abatement programs. In Canada, various programs were introduced after 1991 to actively reduce the artificial sky brightness, notably from street and public lighting (Royal Astronomical Society of Canada, 2011). The worldwide reference for such programs is the International Dark-Sky association (International Dark-Sky Association, 2012).

| | **Bright area change** | | **Urban Population change** | |
|---|---|---|---|---|
| | Absolute ($km^2$) | % | Absolute (people) | % |
| **Russia** | -25'231 | -24.85% | -5'842'556 | -5.35% |
| **Ukraine** | -9'899 | -62.87% | -3'543'432 | -10.16% |
| **Moldova** | -610 | -77.81% | -249'382 | -14.43% |
| **Lithuania** | -84 | -10.41% | -256'730 | -10.28% |
| **Latvia** | -17 | -3.14% | -269'613 | -14.92% |
| **Georgia** | -11 | -2.30% | -332'121 | -12.48% |

**Exhibit14** – List of countries where the decrease of bright areas between 1992 and 2009 coincided with a considerable reduction of the urban population. The urban population figures are estimates from the World Bank (World Bank Group, 2012).

## 4.4 Dynamics of agglomerations

In this final section presenting our results, we will analyze agglomerations instead of countries. The goal is to identify the largest and respectively the most rapidly growing conurbations. We decided, however, not to use the administrative boundaries but to define agglomerations using a threshold method combined with a segmentation function to identify the size of each contiguous cluster. Like in the previous analysis of countries, a binary version of each image is made. All pixels with a digital number above the threshold are given a value of 1, the others, a value of 0. Next, the clusters' sizes are calculated and the name of the biggest and closest city is assigned by visual inspection. The results for different threshold are presented in Exhibit 15 and 16. Clearly, a diverging dynamic is observable between agglomerations in developing countries, which have seen an extraordinary growth, and agglomerations in developed countries, which have seen stagnation and even a decrease in size.

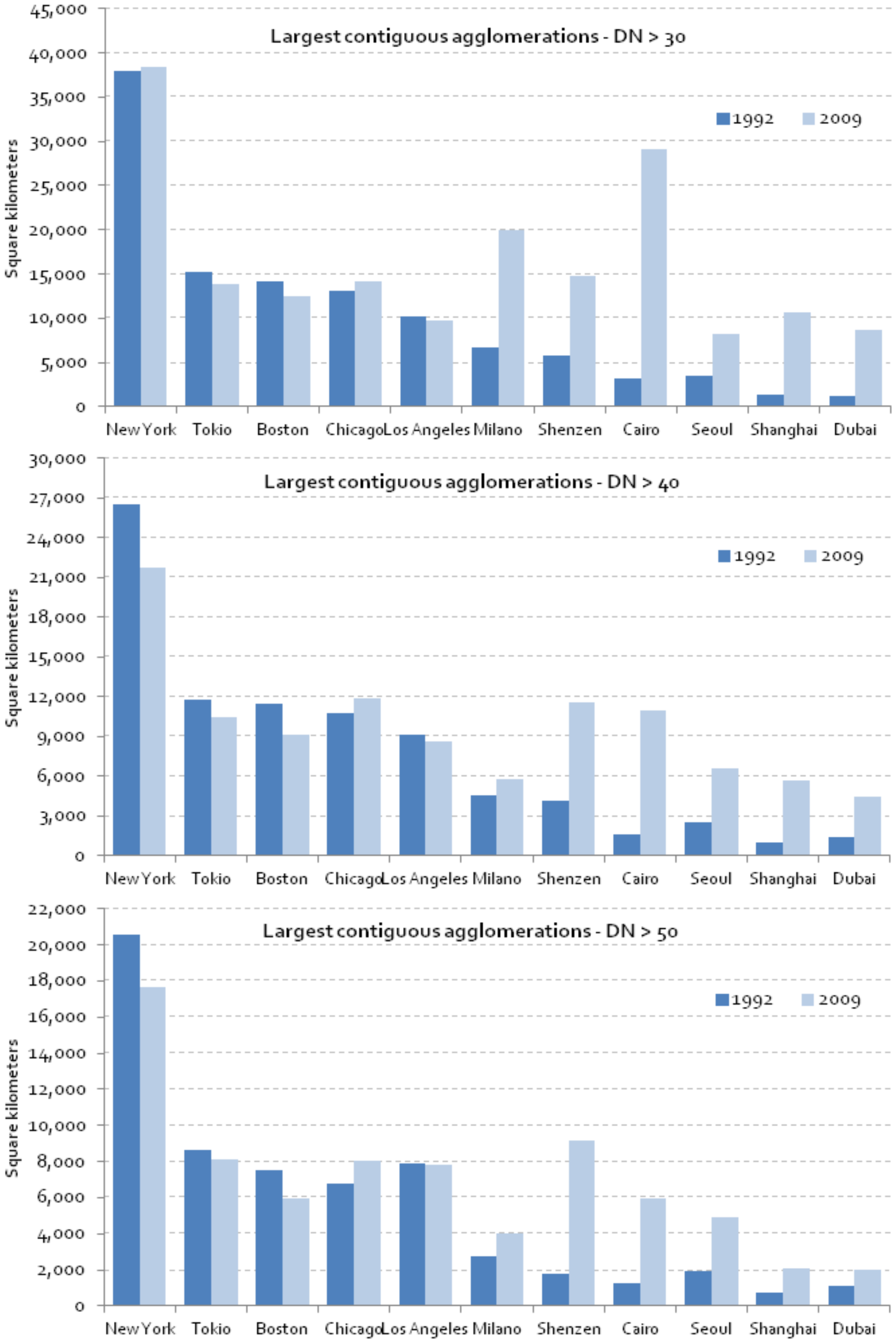

**Exhibit15** – Largest contiguous light agglomerations detected with increasing thresholds: 30 (upper), 40 (middle) and 50 (bottom). The Western cities, in the left part of the graph, show a stable or slightly decreasing trend, whereas the other half experienced an incredible growth. Note how the trends are independent of the chosen threshold. Interesting are the remarkable growth of Cairo, due to the extraordinary development of the Nile Delta, and the agglomeration of Milano, resulting from the coalescence with the surrounding Monza, Bergamo and Brescia.

# 1992

| DN > 30 | | DN > 40 | | DN > 50 | |
|---|---|---|---|---|---|
| Area (km²) | City | Area (km²) | City | Area (km²) | City |
| 37'940 | New York | 26'533 | New York | 20'584 | New York |
| 18'263 | Brussels | 11'753 | Tokyo | 8'640 | Tokyo |
| 15'310 | Tokyo | 11'495 | Boston | 7'895 | Los Angeles |
| 14'101 | Boston | 10'812 | Chicago | 7'514 | Boston |
| 13'791 | Liverpool | 10'738 | Brussels | 6'761 | Chicago |
| 13'069 | Chicago | 9'159 | Toronto | 5'599 | Washington |
| 11'691 | Toronto | 7'452 | Washington | 4'588 | Nagoya |
| 10'249 | Los Angeles | 6'895 | London | 4'513 | Dallas |
| 8'568 | Washington | 6'231 | Nagoya | 4'419 | London |
| 8'541 | London | 5'834 | Dallas | 4'266 | Houston |
| 7'680 | Nagoya | 5'455 | Houston | 3'990 | Osaka |
| 6'999 | Montreal | 5'366 | Osaka | 3'882 | Detroit |
| 6'755 | Osaka | 5'219 | Köln | 3'716 | Toronto |
| 6'639 | Milano | 4'963 | Montreal | 3'700 | Miami |
| 6'473 | Köln | 4'860 | Detroit | 3'587 | Cleveland |
| 6'321 | Dallas | 4'745 | Cleveland | 3'500 | San Francisco |
| 6'098 | Huston | 4'533 | Milano | 3'464 | Atlanta |
| 6'083 | Detroit | 4'330 | San Francisco | 3'232 | Montreal |
| 5'796 | Shenzhen | 4'303 | Tampa | 3'067 | Köln |
| 5'723 | Cleveland | 4'238 | Miami | 2'920 | Paris |

# 2009

| DN > 30 | | DN > 40 | | DN > 50 | |
|---|---|---|---|---|---|
| Area (km²) | City | Area (km²) | City | Area (km²) | City |
| 38'418 | New York | 21'750 | New York | 17'649 | New York |
| 29'171 | Cairo | 11'889 | Chicago | 9'143 | Shenzhen |
| 19'970 | Milano | 11'607 | Shenzhen | 8'113 | Tokyo |
| 15'496 | Brussels | 11'016 | Cairo | 8'049 | Chicago |
| 14'836 | Shenzhen | 10'468 | Tokyo | 7'798 | Los Angeles |
| 14'146 | Chicago | 9'119 | Boston | 5'957 | Cairo |
| 13'810 | Tokyo | 8'672 | Los Angeles | 5'952 | Boston |
| 12'441 | Boston | 7'503 | Tampa | 5'526 | Dallas |
| 10'713 | Shanghai | 7'149 | Brussels | 5'343 | Atlanta |
| 10'225 | Orlando | 6'555 | Seoul | 5'046 | Washington |
| 9'732 | Los Angeles | 6'316 | Washington | 4'998 | Houston |
| 8'613 | Dubai | 6'300 | Atlanta | 4'912 | Seoul |
| 8'213 | Seoul | 6'261 | Dallas | 4'142 | Detroit |
| 7'978 | Atlanta | 5'826 | Huston | 4'088 | Osaka |
| 7'623 | Toronto | 5'799 | Milano | 4'004 | Milano |
| 7'415 | Tel Aviv | 5'709 | Shanghai | 3'914 | Sao Paulo |
| 7'372 | Dallas | 5'440 | Nagoya | 3'900 | Nagoya |
| 7'256 | Moscow | 5'201 | Luxor | 3'836 | London |
| 7'041 | London | 5'185 | Detroit | 3'753 | Miami |
| 6'825 | Huston | 5'176 | London | 3'518 | Luxor |

**Exhibit16** - The 20 largest contiguous light agglomerations in 1992 (upper table) and 2009 (bottom table), with different DN thresholds. The regions were named after the included city with the largest population, but often embrace also other cities. For example, whatever the threshold value, the New York light region also includes Philadelphia, Bridgeport and Hartford. Note in particular the newcomers in 2009 from China (Shenzhen, Shanghai), Korea (Seoul), Egypt (Cairo), and Brazil (Sao Paulo).

## 5. Conclusions

In this research paper, we have analyzed the dynamics, and the spatial distribution of nightlight. We started with an extensive literature review, which made it possible to do a comparative and a qualitative analysis of the broad range of preceding studies. This allowed us to set up a best-practice for the use of nighttime light images in different research projects. Firstly, when covering multiple years of data, we recommend that an inter-calibration of the images should be performed. Then, if not relevant for the study, the gas flares should be removed. Next, a re-projection of the images with an equal-area method should be carried out. And finally, to make reproduction and comparison of different studies possible, each analysis should explicitly mention the version of the Stable Lights product being used.

Using an ArcGIS software package, we first calculated the dynamics of our planet's mean center of light. It was found that, over the past 17 years, this has been gradually shifting eastwards over a distance of roughly 1000 km, at a pace of about 60 km per year. This is less than half of the 140 km per year based on GDP as has been calculated by the McKinsey Global Institute (MGI, 2012).

Then, we introduced the new concept of spatial light Gini coefficients. When removing non-habitable land from the calculations, we came up with the astonishing result that the Lorenz curves and the Gini coefficients converge for all the countries in our analysis. This result suggests that the spatial configuration of settlements follows a universal pattern across different countries, even though the considered areas can differ by up to two orders of magnitude.

When looking at the historical evolution of the spatial light Gini coefficients, we saw a gradual increase over the timespan of our observations, which is the past 17 years. This is indicative of a centralization of light. A detailed analysis revealed that, between 1992 and 2009, the worldwide dimly lit surface grew 49%, whereas the bright area expanded 19%. The total area of dimly lit settlements has, both in relative as in absolute terms, increased more than the area of the bright cities. It may be counter-intuitive, but this is the major driver behind the observed increase in spatial light Gini coefficients over that period.

This research was further completed with a detailed analysis of bright light (city) growth covering 160 different countries. The results showed that nighttime light images provide the perfect tool to monitor the expansion of developing countries (like India and Brazil), the growth of new agglomeration sizes (like Shanghai in China or the Nile delta in Egypt), the regression in countries suffering from demographic decline and a reduction in urban population (like Russia and the Ukraine) and the success of light pollution abatement programs in western countries like Canada and the United Kingdom.

## Acknowledgements

This paper is an outgrowth of the master thesis completed in Nov. 2012 by Nicola Pestalozzi entitled "Earth Nighttime Lights to study economic and urban growth", performed under the supervision of the other authors. Additional material, including images and videos that were made for the master thesis project, can be found on www.worldatnight.ethz.ch. The thesis is available online at www.er.ethz.ch/publications/MAS-Nicola_Pestalozzi_Dec12.pdf.

We wish to heartily thank the whole Land Use Engineering Research Group (LUE) of Professor Hans Rudolf Heinimann. We would like to mention Daniel Trüssel and Monika Niederhuber in particular for sharing their expertise on ArcGIS. Thanks also to Luc Girardin and Sebastian Schutte from the International Conflict Research Group (ICR) of Professor Lars-Erik Cedermann for their help, especially in the initial phase of the work. This work exemplifies the collaborative and collegial atmosphere within the Risk Center at ETH Zurich (www.riskcenter.ethz.ch), which includes the three chairs of Prof. Heiniman, Prof. Cederman and Prof. Sornette.

# 6. Appendix

## 6.1 Evaluation Table: Literature Review

| Application | Publication | Dataset version specified | Single year | Multiple years | Intercalibration | Gas Flares Removed | Equal-area reprojection |
|---|---|---|---|---|---|---|---|
| Economic Activity | (Doll, et al., 2005) | | x | | | | x |
| | (Sutton, et al., 2007) | | | x | | | |
| | (Zhao, et al., 2010) | 2 | | x | x | | x |
| | (Bhandari & Roychowdhury, 2011) | | x | | x | x | |
| | (Ghosh, et al., 2009), | | x | | | | x |
| | (Chen & Nordhaus, 2011) | 4 | | x | x | | |
| | (Kulkarni, et al., 2011) | | | x | | | |
| | (Henderson, et al., 2012) | 4 | | x | | x | |
| Demographics/Politics | (Sutton, 1997) | | x | | | | |
| | (Sutton, et al., 2001) | | x | | | | x |
| | (Lo, 2001) | | x | | | | x |
| | (Zhuo, et al., 2009) | 2 | x | | | | x |
| | (Amaral, et al., 2006) | | | x | | | |
| | (Elvidge, et al., 2009) | | x | | | | |
| | (Imhoff, et al., 1997) | | x | | | | x |
| | (Henderson, et al., 2003). | | x | | | | |
| | (Small, et al., 2005) | | | x | | | |
| | (Elvidge, et al., 2007) | | x | | | x | x |
| | (Lo, 2002) | | x | | | | x |
| | (Ma, et al., 2012) | 4 | | x | x | | |
| | (Agnew, et al., 2008) | 1 | | x | | | x |
| | (Witmet & Loughlin, 2011) | 4 | | x | x | | |
| | (Min, 2009) | | | x | | | |
| Environmental | (Elvidge, et al., 1997) | | x | | | | x |
| | (Kiran Chand, et al., 2007) | | x | | | | |
| | (Amaral, et al., 2005) | | x | | | | |
| | (Letu, et al., 2010) | | x | | | | x |
| | (Kiran Chand, et al., 2007) | | | x | | | |
| | (Fuller & Fulk, 2010) | | x | | | | |
| | (Elvidge, et al., 2010) | | | x | | | |
| | (Kiyofuji & Saitoh, 2004) | | | x | | | |
| | (Aubrecht, et al., 2008) | | x | | | | |
| | (Cinzano, et al. 2000) | | x | | | | x |
| | (Cinzano & Elvidge, 2004) | | x | | | | x |
| | (Tavoosi, et al., 2009) | | | x | | | |
| **TOTAL** | **35** | 7 | 20 | 15 | 5 | 3 | 14 |

**ExhibitA1** – Evaluation of data and methods used in the selected publications. Only a minority of the publications performed an accurate preprocessing of the database. The cells were left blank if the operation was not explicitly mentioned in the study.

## 6.2 Change in bright areas

| Country | 1992 | 2009 | Diff | % |
|---|---|---|---|---|
| China | 42'255 | 137'649 | +95'394 | +226% |
| India | 26'411 | 47'791 | +21'380 | +81% |
| Brazil | 36'694 | 57'926 | +21'232 | +58% |
| Egypt | 12'231 | 32'973 | +20'742 | +170% |
| Italy | 41'167 | 57'662 | +16'495 | +40% |
| Iran | 16'085 | 30'700 | +14'615 | +91% |
| Spain | 25'768 | 40'191 | +14'423 | +56% |
| Saudi Arabia | 20'230 | 31'901 | +11'671 | +58% |
| South Korea | 9'173 | 20'389 | +11'216 | +122% |
| Mexico | 25'726 | 34'750 | +9'024 | +35% |
| Argentina | 11'607 | 19'906 | +8'299 | +71% |
| Thailand | 5'320 | 12'682 | +7'362 | +138% |
| Malaysia | 3'120 | 10'422 | +7'302 | +234% |
| Portugal | 4'208 | 10'709 | +6'501 | +154% |
| France | 35'514 | 41'569 | +6'055 | +17% |
| Poland | 9'978 | 15'713 | +5'735 | +57% |
| United Arab Emirates | 4'767 | 9'657 | +4'890 | +103% |
| Turkey | 6'896 | 11'617 | +4'721 | +68% |
| Vietnam | 417 | 4'784 | +4'367 | +1'047% |
| Indonesia | 5'214 | 9'152 | +3'938 | +76% |
| Libya | 2'748 | 6'272 | +3'524 | +128% |
| Israel | 4'971 | 8'026 | +3'055 | +61% |
| Oman | 1'757 | 4'548 | +2'791 | +159% |
| Algeria | 4'296 | 6'934 | +2'638 | +61% |
| Chile | 2'664 | 5'128 | +2'464 | +92% |
| Greece | 2'836 | 5'215 | +2'379 | +84% |
| Taiwan | 6'749 | 8'802 | +2'053 | +30% |
| Romania | 1'334 | 3'287 | +1'953 | +146% |
| Croatia | 651 | 2'420 | +1'769 | +272% |
| Morocco | 2'350 | 4'113 | +1'763 | +75% |
| Jordan | 1'126 | 2'867 | +1'741 | +155% |
| Ecuador | 1'606 | 3'230 | +1'624 | +101% |
| Peru | 1'823 | 3'436 | +1'613 | +88% |
| Colombia | 5'769 | 7'303 | +1'534 | +27% |
| Pakistan | 8'411 | 9'825 | +1'414 | +17% |
| Australia | 12'303 | 13'696 | +1'393 | +11% |
| Switzerland | 4'121 | 5'468 | +1'347 | +33% |
| Kuwait | 1'180 | 2'411 | +1'231 | +104% |
| Syria | 2'241 | 3'389 | +1'148 | +51% |
| Austria | 2'439 | 3'568 | +1'129 | +46% |
| Serbia | 1'268 | 2'373 | +1'105 | +87% |
| Qatar | 928 | 2'005 | +1'077 | +116% |
| Venezuela | 10'770 | 11'800 | +1'030 | +10% |
| Philippines | 1'464 | 2'480 | +1'016 | +69% |
| Angola | 228 | 1'211 | +983 | +431% |
| Trinidad and Tobago | 639 | 1'620 | +981 | +154% |
| Yemen | 577 | 1'483 | +906 | +157% |
| Finland | 11'059 | 11'937 | +878 | +8% |
| Ireland | 1'706 | 2'493 | +787 | +46% |
| Tunisia | 1'918 | 2'613 | +695 | +36% |
| Bolivia | 861 | 1'547 | +686 | +80% |
| Lebanon | 530 | 1'203 | +673 | +127% |
| Bosnia | 11 | 639 | +628 | +5'709% |
| Paraguay | 759 | 1'379 | +620 | +82% |
| Sudan | 920 | 1'508 | +588 | +64% |
| Cuba | 510 | 1'060 | +550 | +108% |
| South Africa | 13'297 | 13'834 | +537 | +4% |
| Hungary | 2'092 | 2'610 | +518 | +25% |
| Slovenia | 387 | 893 | +506 | +131% |
| Uruguay | 957 | 1'438 | +481 | +50% |
| Guatemala | 553 | 1'014 | +461 | +83% |
| Cyprus | 634 | 1'094 | +460 | +73% |
| Czech Republic | 4'827 | 5'249 | +422 | +9% |
| Turkmenistan | 1'091 | 1'498 | +407 | +37% |
| Myanmar | 268 | 673 | +405 | +151% |
| Honduras | 396 | 792 | +396 | +100% |
| Estonia | 687 | 1'073 | +386 | +56% |
| Ghana | 688 | 1'072 | +384 | +56% |
| Mozambique | 170 | 532 | +362 | +213% |
| Dominican Republic | 796 | 1'151 | +355 | +45% |
| Afghanistan | 189 | 499 | +310 | +164% |
| Brunei | 236 | 546 | +310 | +131% |
| Albania | 16 | 291 | +275 | +1'719% |
| Panama | 520 | 788 | +268 | +52% |
| Sri Lanka | 517 | 758 | +241 | +47% |
| Norway | 7'671 | 7'903 | +232 | +3% |
| Senegal | 186 | 396 | +210 | +113% |
| Cote d'Ivoire | 400 | 605 | +205 | +51% |
| El Salvador | 427 | 632 | +205 | +48% |
| Cambodia | 33 | 209 | +176 | +533% |

| Country | 1992 | 2009 | Diff | % |
|---|---|---|---|---|
| Nigeria | 2'978 | 3'173 | +195 | +7% |
| Laos | 60 | 239 | +179 | +298% |
| Macedonia | 250 | 422 | +172 | +69% |
| Ethiopia | 151 | 301 | +150 | +99% |
| Iraq | 4'531 | 4'680 | +149 | +3% |
| Nicaragua | 295 | 441 | +146 | +49% |
| Burkina Faso | 118 | 256 | +138 | +117% |
| Botswana | 158 | 295 | +137 | +87% |
| Armenia | 204 | 338 | +134 | +66% |
| Mali | 121 | 254 | +133 | +110% |
| Bulgaria | 1'213 | 1'339 | +126 | +10% |
| Benin | 86 | 194 | +108 | +126% |
| Costa Rica | 923 | 1'027 | +104 | +11% |
| Kosovo | 90 | 194 | +104 | +116% |
| Zambia | 592 | 695 | +103 | +17% |
| Malawi | 186 | 285 | +99 | +53% |
| Montenegro | 87 | 180 | +93 | +107% |
| Mauritania | 60 | 150 | +90 | +150% |
| Suriname | 112 | 196 | +84 | +75% |
| Uganda | 97 | 178 | +81 | +84% |
| Swaziland | 67 | 142 | +75 | +112% |
| Mongolia | 245 | 311 | +66 | +27% |
| Lesotho | 34 | 92 | +58 | +171% |
| Namibia | 235 | 287 | +52 | +22% |
| Guinea | 54 | 104 | +50 | +93% |
| Tanzania | 297 | 345 | +48 | +16% |
| Cameroon | 291 | 337 | +46 | +16% |
| Gabon | 169 | 215 | +46 | +27% |
| The Gambia | 21 | 63 | +42 | +200% |
| Bahamas | 221 | 262 | +41 | +19% |
| Belize | 32 | 70 | +38 | +119% |
| Haiti | 68 | 97 | +29 | +43% |
| Madagascar | 72 | 98 | +26 | +36% |
| Chad | 39 | 63 | +24 | +62% |
| Togo | 118 | 138 | +20 | +17% |
| Djibouti | 14 | 32 | +18 | +129% |
| Sierra Leone | 10 | 25 | +15 | +150% |
| Guyana | 34 | 47 | +13 | +38% |
| Timor Leste | 3 | 16 | +13 | +433% |
| Niger | 106 | 118 | +12 | +11% |
| Eritrea | 33 | 41 | +8 | +24% |
| Vanuatu | 9 | 14 | +5 | +56% |
| Congo | 247 | 249 | +2 | +1% |
| Fiji | 26 | 24 | -2 | -8% |
| Guinea-Bissau | 7 | 0 | -7 | -100% |
| New Zealand | 1'975 | 1'966 | -9 | -1% |
| Rwanda | 70 | 61 | -9 | -13% |
| Burundi | 44 | 34 | -10 | -23% |
| Georgia | 478 | 467 | -11 | -2% |
| Latvia | 542 | 525 | -17 | -3% |
| Central African | 33 | 14 | -19 | -58% |
| Nepal | 146 | 126 | -20 | -14% |
| North Korea | 95 | 66 | -29 | -31% |
| Kenya | 467 | 437 | -30 | -6% |
| Jamaica | 564 | 518 | -46 | -8% |
| Papua New | 183 | 123 | -60 | -33% |
| Congo, DRC | 557 | 479 | -78 | -14% |
| Lithuania | 807 | 723 | -84 | -10% |
| Kyrgyzstan | 559 | 419 | -140 | -25% |
| Netherlands | 11'250 | 11'046 | -204 | -2% |
| Iceland | 616 | 397 | -219 | -36% |
| Bangladesh | 1'254 | 1'012 | -242 | -19% |
| Azerbaijan | 1'306 | 1'009 | -297 | -23% |
| Zimbabwe | 868 | 494 | -374 | -43% |
| Tajikistan | 634 | 149 | -485 | -76% |
| Belarus | 3'052 | 2'541 | -511 | -17% |
| Moldova | 784 | 174 | -610 | -78% |
| Germany | 34'373 | 33'505 | -868 | -3% |
| Denmark | 3'170 | 1'901 | -1'269 | -40% |
| Slovakia | 2'684 | 1'256 | -1'428 | -53% |
| Kazakhstan | 6'811 | 5'366 | -1'445 | -21% |
| Japan | 59'241 | 57'409 | -1'832 | -3% |
| Belgium | 14'195 | 12'166 | -2'029 | -14% |
| Uzbekistan | 5'603 | 2'606 | -2'997 | -53% |
| Sweden | 14'654 | 9'020 | -5'634 | -38% |
| United States | 438'324 | 432'008 | -6'316 | -1% |
| United Kingdom | 43'746 | 36'220 | -7'526 | -17% |
| Ukraine | 15'745 | 5'846 | -9'899 | -63% |
| Russia | 101'529 | 76'298 | -25'231 | -25% |
| Canada | 79'571 | 52'176 | -27'395 | -34% |

**ExhibitA2** – Change in bright areas for all the countries that have been analyzed, in decreasing order from the largest absolute increase. A bright lit area has pixels with a DN>30.

## 6.3 The Nile delta, Shanghai and Shenzhen-Guangzhou

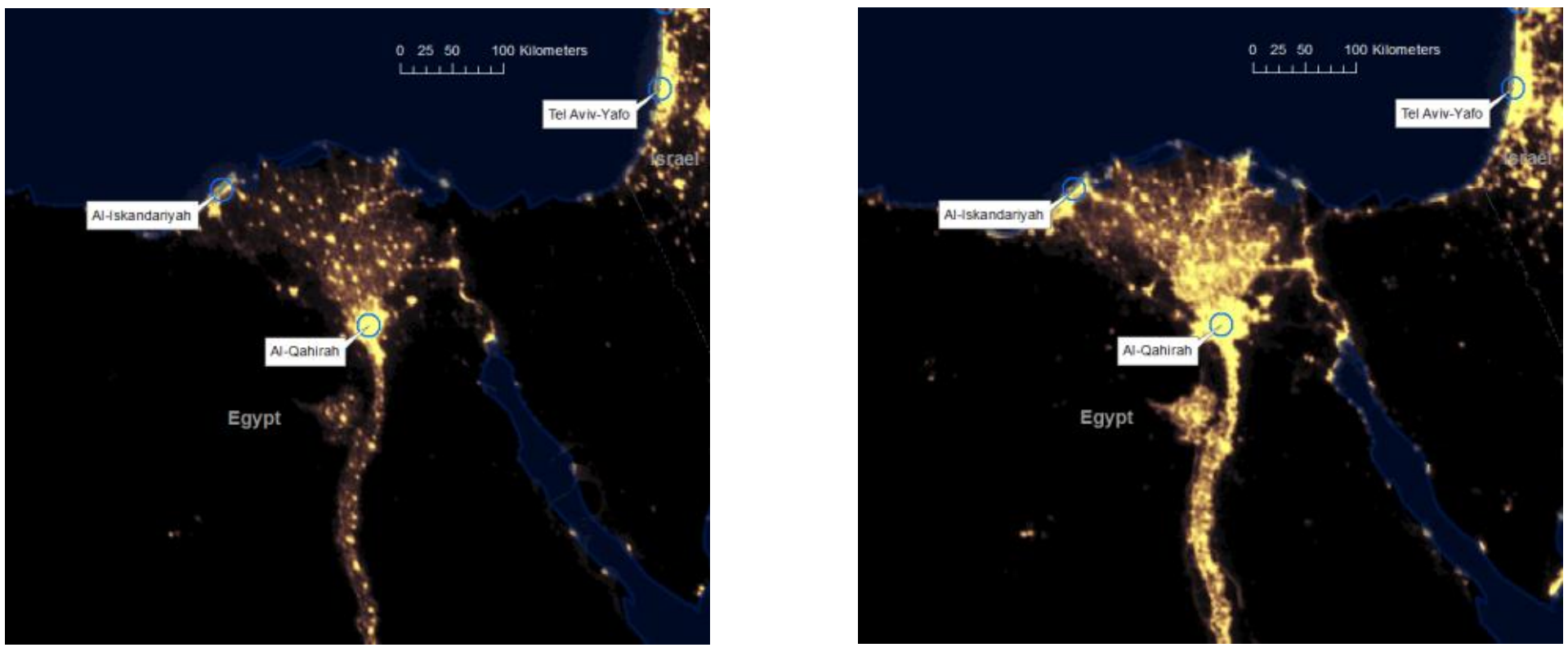

**ExhibitA3** – Nile delta in 1992 (left) and 2009 (right). Nearly the totality of light emission of Egypt comes from this area and the coast of the Nile. Note the consolidations along the interconnections between the settlements.

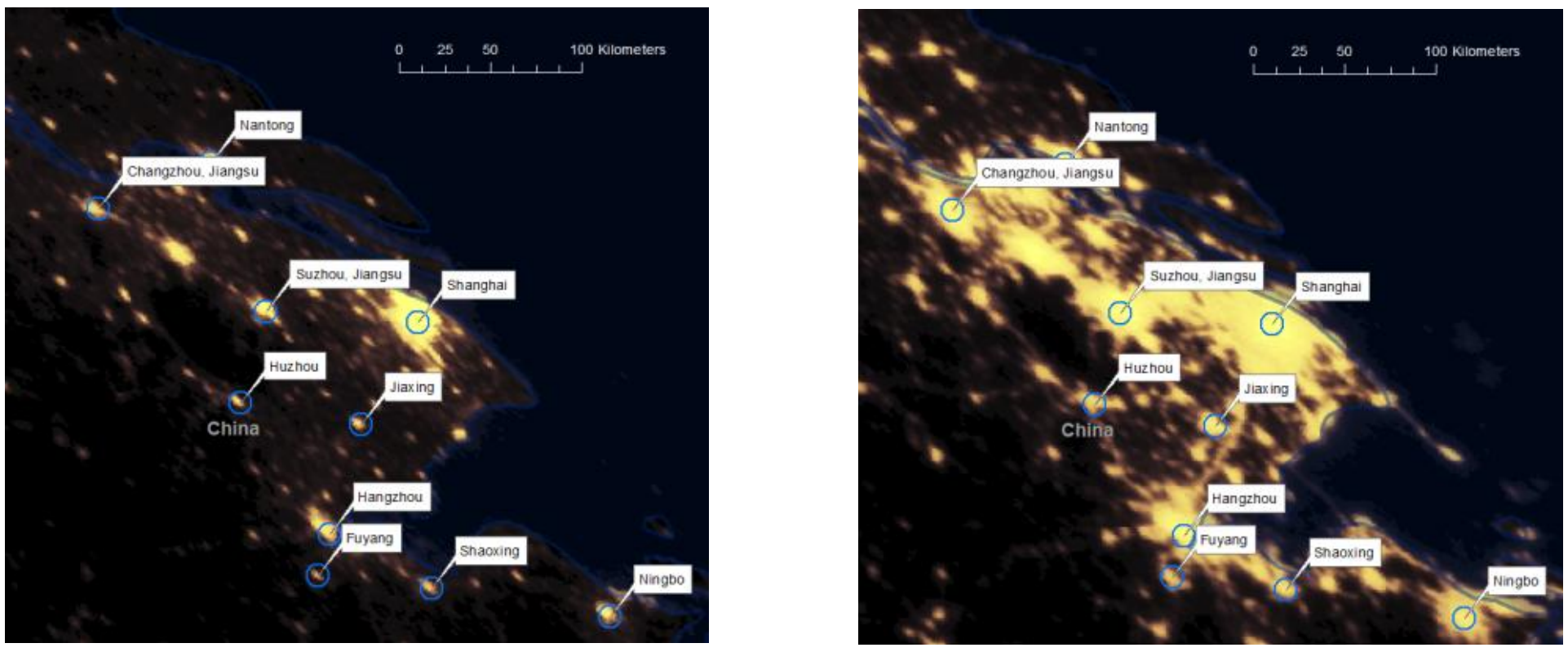

**ExhibitA4** – Impressive growth of lights in China around Shanghai from 1992 (left) to 2009 (right).

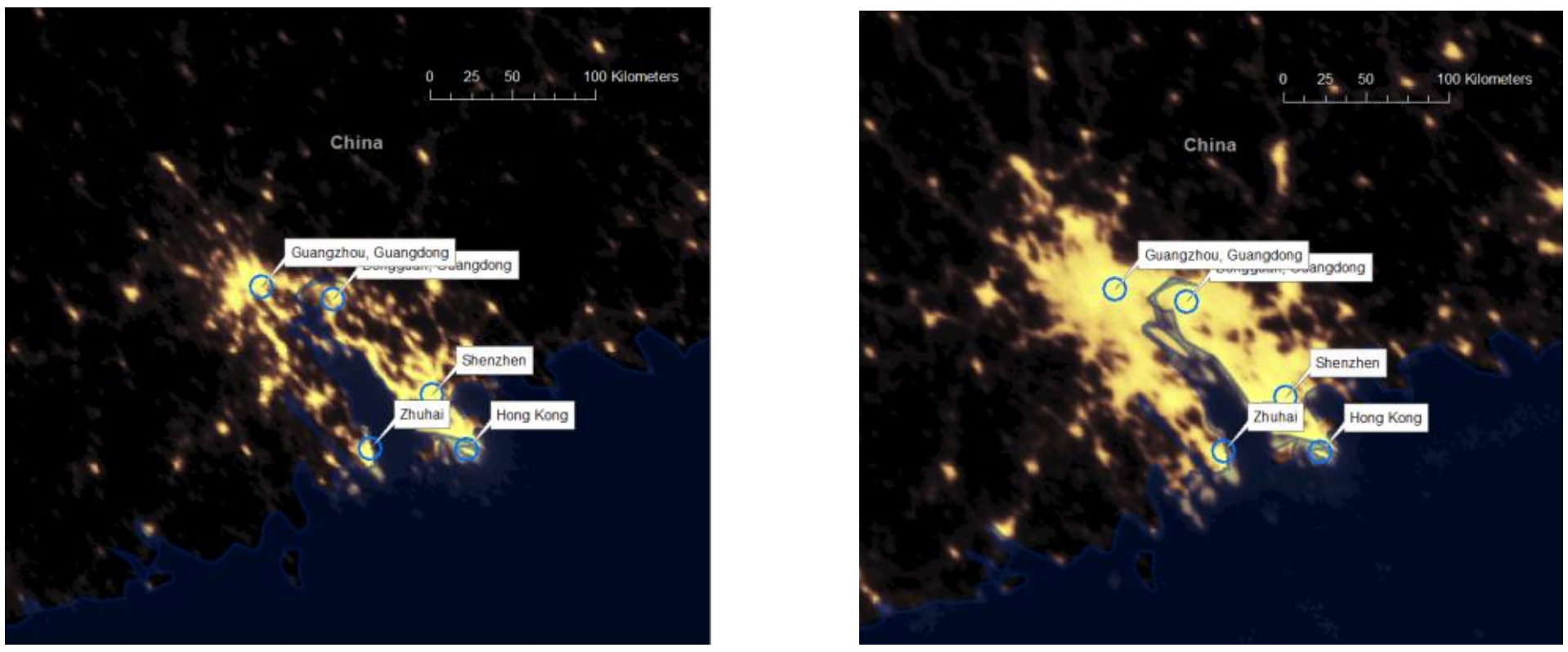

**ExhibitA5** – Lights growth for the agglomeration of Shenzhen-Guangzhou from 1992 (left) to 2009 (right).